\RequirePackage{lineno}
\documentclass[twocolumn,nofootinbib,floatfix]{revtex4-1}

\usepackage{amsmath,amssymb}
\usepackage{cancel}
\usepackage{graphicx}
\usepackage{epsfig}
\usepackage{amsmath,amsfonts,amssymb}
\usepackage{colortbl}
\usepackage{hhline}
\usepackage{pifont}
\usepackage{multirow}

\newcommand{\mj}{m_{J}}
\newcommand{\ptj}{p_{TJ}}

\begin{document}

\newcommand{\Gent}{{\tt GenT}$^3$}
\newcommand{\gentx}{{\tt GenT}$^x$}

\title{Taming modeling uncertainties with Mass Unspecific Supervised Tagging}

\author{J.~A.~Aguilar-Saavedra}
\affiliation{Instituto de F\'{\i}sica Te\'{o}rica UAM-CSIC, Campus de Cantoblanco, E-28049 Madrid, Spain}

\begin{abstract}
We address the modeling dependence of jet taggers built using the method of Mass Unspecific Supervised Tagging, by using two different parton showering and hadronisation schemes.
We find that the modeling dependence of the results --- estimated by using different schemes in the design of the taggers and applying them to the same type of data --- is rather small, even if the jet substructure varies significantly between the two schemes. These results add great value to the use of generic supervised taggers for new physics searches.
\end{abstract}

\maketitle

\tableofcontents

\section{Introduction}

Elusive signals of new particles involving boosted hadronic jets may arise in a variety of models, see for example Refs.~\cite{Aguilar-Saavedra:2015iew,Agashe:2016kfr,Agashe:2017wss,Aguilar-Saavedra:2019adu,Aguilar-Saavedra:2021smc}. Their detection at the Large Hadron Collider (LHC) is quite demanding and requires tools that distinguish between `signal' jets originating from boosted massive particles, and `background' jets from quarks and gluons produced by QCD interactions. Jet substructure tools~\cite{Butterworth:2008iy,Thaler:2010tr,Thaler:2011gf,Larkoski:2014gra,Moult:2016cvt} originally introduced for the discrimination of Standard Model (SM) heavy particles (top quarks and $W/Z/H$ bosons) from QCD jets, are essential for this goal.

In order not to rely on specific assumptions about the nature of the new particles, a generic tagger is necessary to search for these elusive signals. This was indeed the motivation for the anti-QCD tagger in Ref.~\cite{Aguilar-Saavedra:2017rzt}. This tagger uses a relatively simple neural network (NN) architecture and is trained with QCD jets (background) and various types of multi-pronged signal jets. It is remarkable that, despite being a fully-supervised tool, the anti-QCD tagger is able to recognise as signal a wide variety of massive multi-pronged jets. The key aspect to achieve this, is the use of the so-called model-independent (MI) data in the training: massive jets with $n= 2,3,4$ prongs (this number can of course be increased above $n=4$) but otherwise phase-space agnostic. As it was shown in Ref.~\cite{Aguilar-Saavedra:2017rzt}, this setup provides sensitivity to six-pronged jets not used in the training. Mass Unspecific Supervised Tagging (MUST)~\cite{Aguilar-Saavedra:2020uhm} extends this concept by including the jet mass ($m_J$) and transverse momentum ($\ptj$) as training variables, so that the taggers are applicable across a quite wide range of $\mj$ and $\ptj$. An alternative to MI data is explored in Ref.~\cite{Cheng:2022gma}.

Generic taggers for multi-pronged jets can also be built by using representation learning, e.g. with an autoencoder~\cite{Heimel:2018mkt,Farina:2018fyg,Cheng:2020dal,Dillon:2021nxw,Atkinson:2021nlt}. Without the need of any signal assumption, but only using background (pseudo-)data, an unsupervised tagger can learn the background features in order to pinpoint outliers, i.e. signal jets that deviate from the known pattern. A great advantage of unsupervised learning is that the tool does not depend on our modeling of the signals and backgrounds, and can directly be trained on data. On the other hand, the performance is worse than with supervised learning, as shown for example in Ref.~\cite{Aguilar-Saavedra:2021utu}.

The application of supervised taggers to data raises concerns about their dependence on the modeling of parton showers and hadronisation, as well as other effects that are not described from first principles but phenomenologically. These issues were studied in Ref.~\cite{Barnard:2016qma} for a $W$ boson tagger using jet images. For that specific setup, a variation of signal efficiency $\varepsilon_\text{sig} = 0.4-0.56$ is found at fixed background rejection $\varepsilon_\text{bkg}^{-1} = 10$ by using several Monte Carlo codes.\footnote{Ref.~\cite{Barnard:2016qma} quotes a variation of up to 50\% in background rejection for fixed signal efficiency. However, anomaly detection tools often use the jet mass a discriminant, using a mass-decorrelated tagging with fixed background rejection. Thus, the variation of the signal efficiency at fixed background rejection is more adequate for the assessment of the dependence on the Monte Carlo setup.} We note that the size of this variation is expected to depend on
\begin{itemize}
\item[(a)] the type of signal jet (mass and prongness);
\item[(b)] the transverse momentum;
\item[(c)] the specific method used to build the tagger: jet images versus jet substructure variables, NN architecture, etc. 
\end{itemize}
In this respect, it is important to point out that Ref.~\cite{Barnard:2016qma}  also finds a variation of the same size, $\varepsilon_\text{sig} = 0.35-0.52$ for $\varepsilon_\text{bkg}^{-1} = 10$, by using as discriminant the jet mass and subjettiness ratio $\tau_{21}$~\cite{Thaler:2010tr,Thaler:2011gf} and pseudo-data generated by several Monte Carlo codes. Obviously, the variation in the latter case is not due to the design of the discriminant --- in other words, whether it is trained using certain Monte Carlo code or another --- but rather to {\it how pseudo-data is}. 

In this paper we address the modeling dependence for a generic tagger built upon MUST. We consider two Monte Carlo hadronisation schemes, using {\scshape Pythia}~\cite{Sjostrand:2007gs} and {\scshape Herwig}~\cite{Bellm:2015jjp}, and explore the differences when training taggers and applying them to pseudo-data obtained with each of these generators. We study 18 benchmarks with signal jets of different mass, transverse momentum and prongness. The conclusions, which we anticipate here, are:
\begin{itemize}
\item[(i)] The dependence of the results on the generator used for the design (training) of the tagger is small, and insignificant in many cases.
\item[(ii)] On the other hand, there is a significant dependence of the results on {\em how pseudo-data is}. 
\end{itemize}
From the above, one can also expect that when applied to real data, the performance will mostly depend on how real data actually is, {\em i.e.} if the subjets are more or less resolved. And of course, this statement not only applies to supervised taggers, but also to unsupervised ones.

The remainder of this paper is structured as follows. In section~\ref{sec:2} we describe our setup for the Monte Carlo generation and training of the NNs. We compare the jet substructure observables obtained with either simulation scheme in section~\ref{sec:3}. When possible, we compare our qualitative results obtained with subjettiness variables with the findings of Ref.~\cite{Barnard:2016qma} using jet images. We compare the tagging performances in section~\ref{sec:4}. Finally, our results are discussed in section~\ref{sec:5}.

\section{Monte Carlo generation and design of the taggers}
\label{sec:2}

In Ref.~\cite{Aguilar-Saavedra:2020uhm} we developed a supervised generic jet tagger, dubbed as {\tt GenT}, in order to discriminate between quark/gluon one-pronged jets and multi-pronged jets from boosted massive particles. Here we train taggers \gentx\ with the same NN architecture, in the transverse momentum range $\ptj \in [200,2200]$ GeV and extending the mass range to $m_J \in[10,500]$ GeV.

QCD jets are generated with {\scshape MadGraph}~\cite{Alwall:2014hca}, in the inclusive process $pp \to jj$. Event samples are generated in 100 GeV bins of $p_T$, starting at $[200,300]$ GeV and up to $p_T \geq 2.2$ TeV. Large event samples are required in order to have sufficient events at high $m_J$: $10^6$ events are generated in each bin of $\ptj$, and both jets are used in the analysis, amounting to a total of 42 million QCD jets, which are used in the training and validation of the NNs, as well as for tests.

The MI data used to train and validate the NNs are generated with {\scshape Protos}~\cite{protos} in the process $pp \to ZS$, with $Z \to \nu \nu$ and $S$ a scalar. We consider the six decay modes
\begin{align}
&  \text{4-pronged (4P):} && S \to u \bar u u \bar u \,,~ S \to b \bar{b} b \bar{b} \,, \notag \\
& \text{3-pronged (3P):} && S \to F \,\nu \,; \quad F \to u d d \,,~ F \to u d b \,, \notag \\
& \text{2-pronged (2P):} && S \to u \bar u \,,~ S \to b \bar b \,,
\label{ec:MIdata}
\end{align}
to generate multi-pronged jets ($F$ is a colour-singlet fermion). To remain as model-agnostic as possible, the $S$ and $F$ decays are implemented with a flat matrix element, so that the decay weight of the different kinematical configurations only corresponds to the four-, three- or two-body phase space. Signal jet samples are also generated in 100 GeV bins of $p_T$. To cover different jet masses, the mass of $S$ (and of $F$ for 3-pronged decays) is randomly chosen event by event within the interval $[10,800]$~GeV, and setting an upper limit $M_S \leq p_T R/2$ to ensure that all decay products are contained in a jet of radius $R=0.8$.

The signals used to evaluate the performance of the taggers are generated with {\scshape MadGraph}. The models of Refs.~\cite{Aguilar-Saavedra:2019adu,Aguilar-Saavedra:2021smc} are implemented in {\scshape Feynrules} ~\cite{Alloul:2013bka} and interfaced to {\scshape MadGraph} using the universal Feynrules output~\cite{Degrande:2011ua}. We consider the production of a neutral gauge boson $Z'$ with decay into SM particles as well as new scalars. We generically denote by $A$ the new scalars decaying into a quark pair, and by $S$ the new scalars decaying into an $AA$ pair. (The actual scalar decays are $Z' \to A_i A_j$, $Z' \to S_i S_j$, with $i\neq j$, but we omit subindices for simplicity and consider $M_{A_i} = M_{A_j}$, $M_{S_i} = M_{S_j}$). The various processes considered are
\begin{itemize}
\item[(1)] $Z' \to WW$, $W \to q \bar q$.
\item[(2)] $Z' \to AA$, $A \to b \bar b$.
\item[(3)] $Z' \to t \bar t$, $t \to W b \to q \bar q b$.
\item[(4)] $Z' \to SS$, with $S \to WW \to 4q$. 
\item[(5)] $Z' \to SS$, with $S \to AA \to 4b$.
\end{itemize}
All signal samples contain $2 \times 10^{5}$ events except the $t \bar t$ samples, which contain $6 \times 10^{5}$ events.

The parton-level event samples are hadronised either with {\scshape Pythia}  8.3 or with {\scshape Herwig} 7.2, with standard settings. The former uses dipole showers by default~\cite{Sjostrand:2004ef}, while the latter uses angular-ordered showers~\cite{Gieseke:2003rz}. In both cases, a fast detector simulation is performed with {\scshape Delphes}~\cite{deFavereau:2013fsa}, using the CMS card. Jets are reconstructed with {\scshape FastJet}~\cite{Cacciari:2011ma} applying the anti-$k_T$ algorithm~\cite{Cacciari:2008gp} with $R=0.8$, and groomed with Recursive Soft Drop~\cite{Dreyer:2018tjj}. 
Jet substructure is characterised by a set of subjettiness variables proposed in~\cite{Thaler:2010tr,Datta:2017rhs},
 \begin{equation}
 \left\{ \tau_1^{(1/2)}, \tau_1^{(1)}, \tau_1^{(2)}, \dots , \tau_{5}^{(1/2)}, \tau_{5}^{(1)}, \tau_{5}^{(2)}, \tau_{6}^{(1)}, \tau_{6}^{(2)} \right\} \,,
 \label{ec:taulist}
 \end{equation}
computed for ungroomed jets.\footnote{One might consider that calculating $\tau_n^{(i)}$ for groomed jets would decrease the dependence on the details of showering and hadronisation, but unfortunately the groomed $\tau_n^{(i)}$ have much less discriminating power~\cite{Aguilar-Saavedra:2020uhm}.} These 17 subjettiness variables constitute the input to the NN together with the groomed jet mass and $p_T$.

Two taggers are built exactly in the same way, but using either {\scshape Pythia} or {\scshape Herwig} showering and hadronisation. The training sets are obtained by dividing the $m_J$ range in ten bins, all of 50 GeV except the first one $[10,50]$ GeV, and the $p_T$ range in 100 GeV bins, starting at $[200,300]$ GeV and up to $[2100,2200]$ GeV. In the lower $p_T$ samples the higher mass bins are dropped, considering the full $m_J$ range only for the $p_T$ bins above 1200 GeV. In each two-dimensional bin of $\mj$ and $\ptj$ we select for the training set 3000 events from each of the six types of signal jets in (\ref{ec:MIdata}), and 18000 background events, in order to have a balanced sample. The proportion of quark and gluon jets in the background samples is $p_T$-dependent. The total size of the training sets is around 5.5 million events. The validation sets used to monitor the NN performance are similar to the training ones.

The NNs are implemented using {\scshape Keras}~\cite{keras} with a {\scshape TensorFlow} backend~\cite{tensorflow}. For the training, a standardisation of the 19 inputs, based on the SM background distributions, is performed. The NNs contain two hidden layers of 2048 and 128 nodes, with Rectified Linear Unit (ReLU) activation for the hidden layers. For the output layer a sigmoid function is used, yielding the NN score, i.e. the signal probability, that can be used to discriminate signal jets from QCD jets. 
The NNs are optimised by minimising the binary cross-entropy loss function, using the Adam~\cite{Adam} algorithm, and a batch size of 64.
The NNs obtained after the training with these large event sets are remarkably stable. We train five instances of each NN and select the ones that give the largest area under the $(\varepsilon_\text{sig},\varepsilon_\text{bkg})$ curve (AUC) for the validation sets.

\section{Substructure observables}
\label{sec:3}

With the default options for parton showering and hadronisation used here, {\scshape Pythia} and {\scshape Herwig} produce the two most differing results among five combinations studied in Ref.~\cite{Barnard:2016qma}. It is very difficult, if not impossible, to fully understand the results obtained next section from the behaviour observed in substructure observables: there are many non-trivial correlations among the inputs to the NNs. However, there a few qualitative aspects that can be learnt by the comparison of the subjettiness variables obtained after {\scshape Pythia} and {\scshape Herwig} simulation.

\begin{figure*}[p]
\begin{center}
\begin{tabular}{cccc}
\includegraphics[width=5.2cm,clip=]{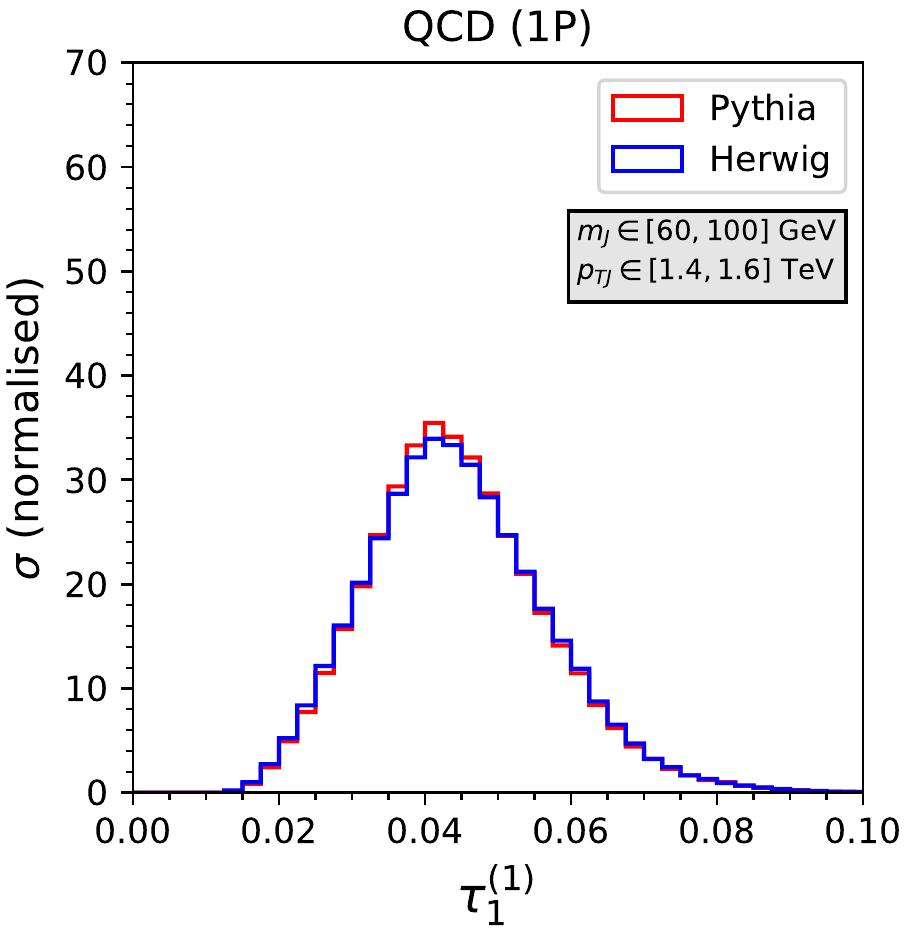} &
\includegraphics[width=5.2cm,clip=]{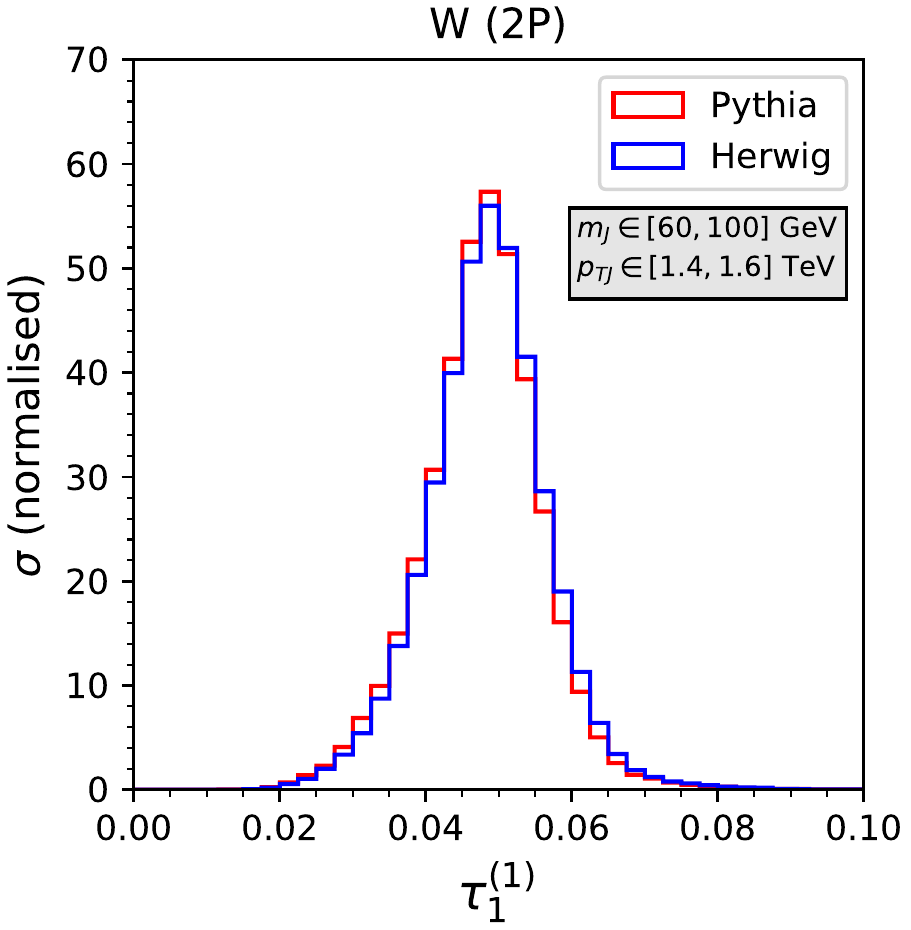} & 
\includegraphics[width=5.2cm,clip=]{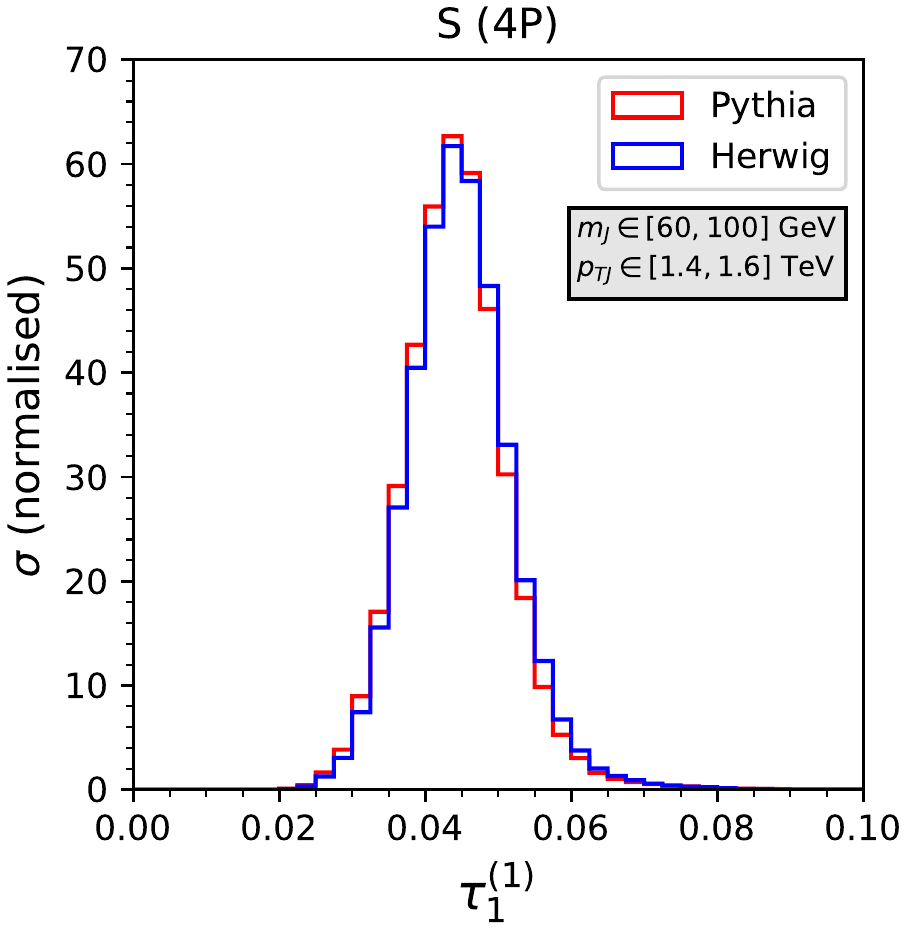} \\
\includegraphics[width=5.2cm,clip=]{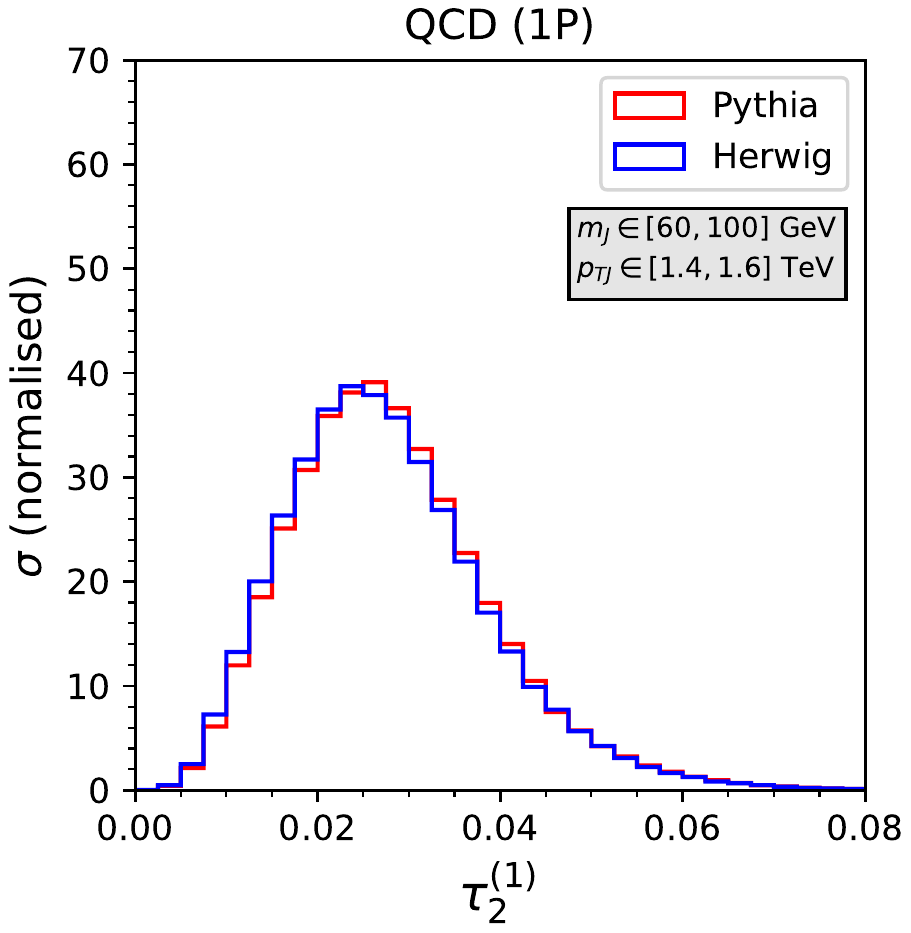} &
\includegraphics[width=5.2cm,clip=]{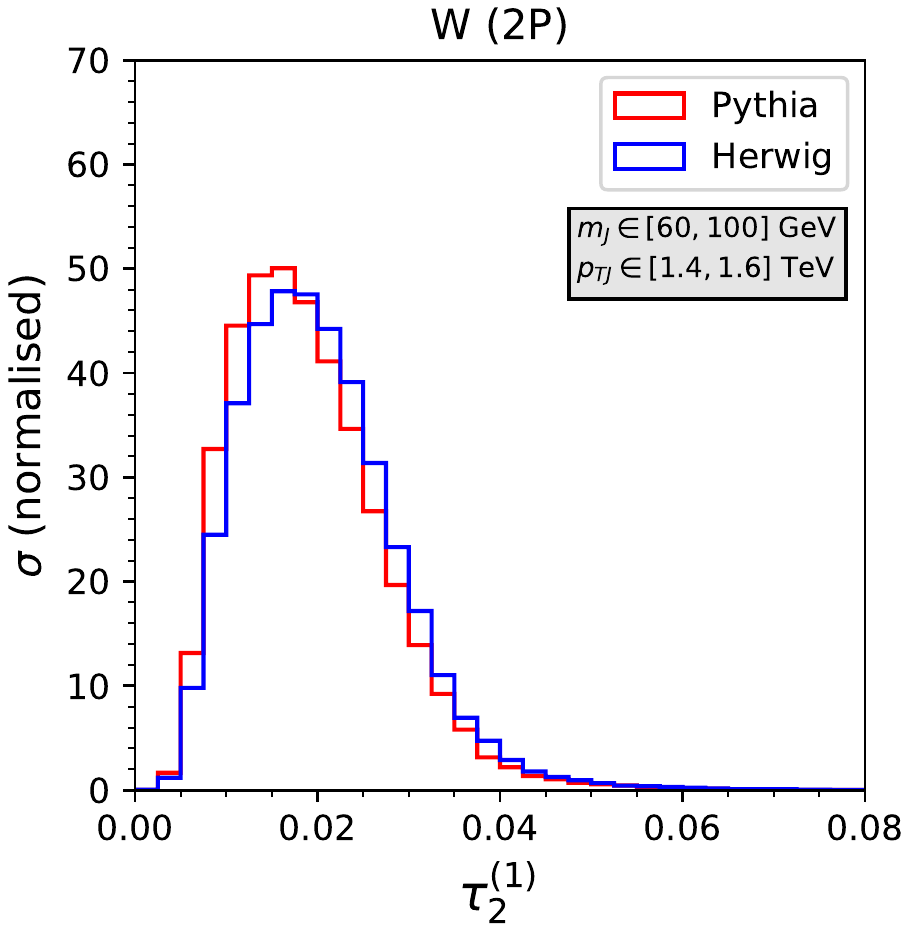} & 
\includegraphics[width=5.2cm,clip=]{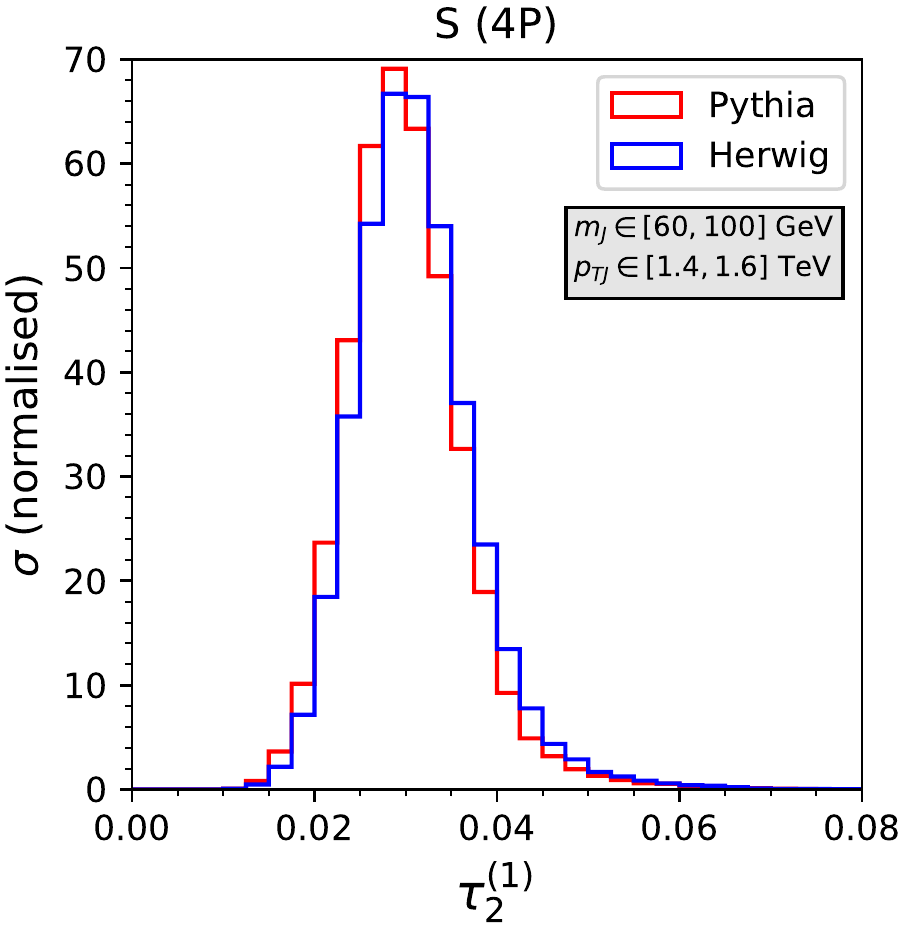} \\
\includegraphics[width=5.2cm,clip=]{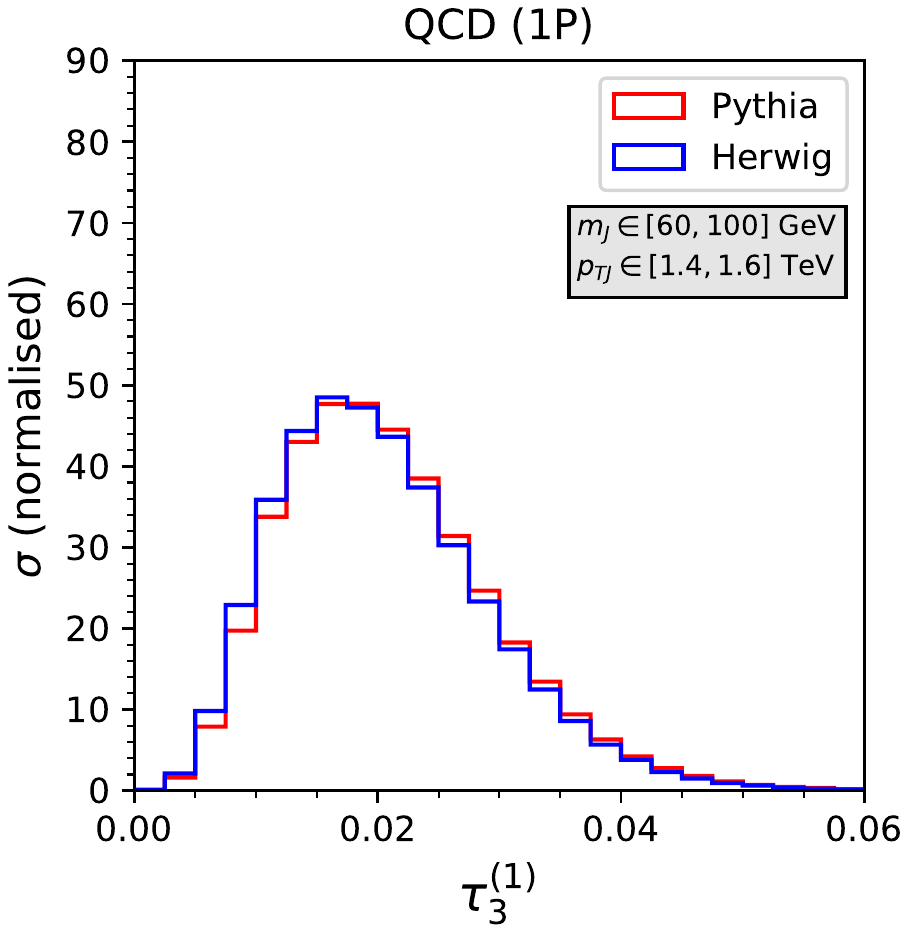} &
\includegraphics[width=5.2cm,clip=]{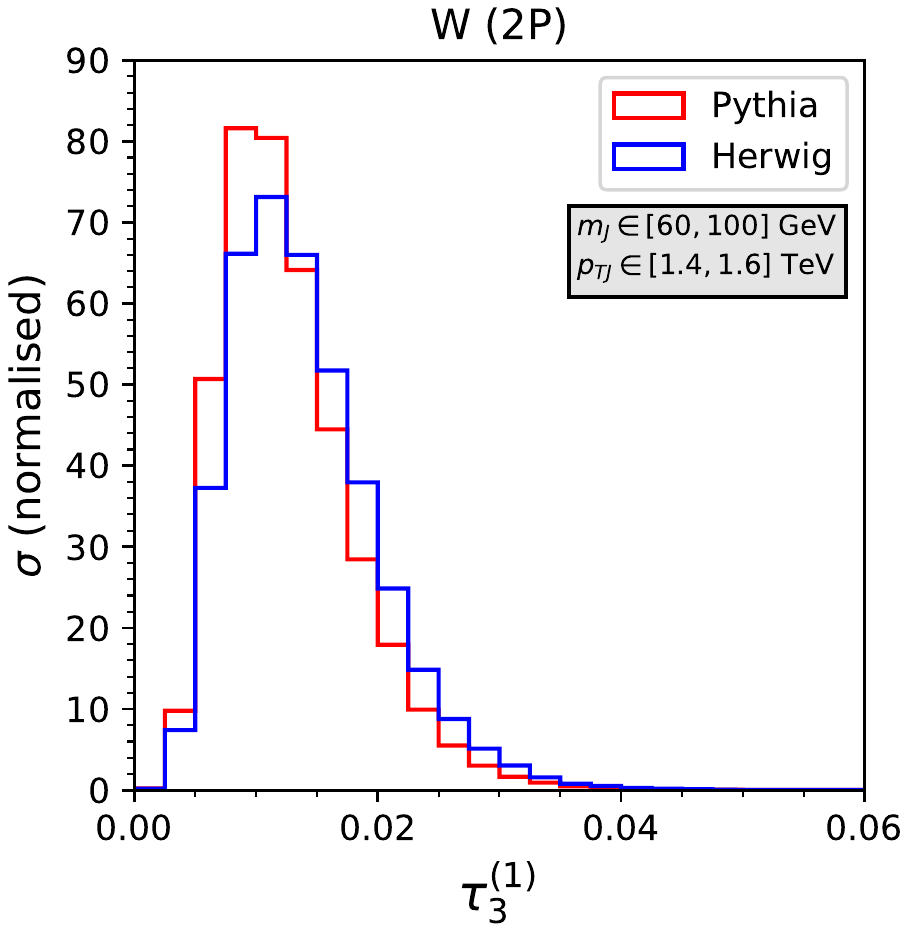} & 
\includegraphics[width=5.2cm,clip=]{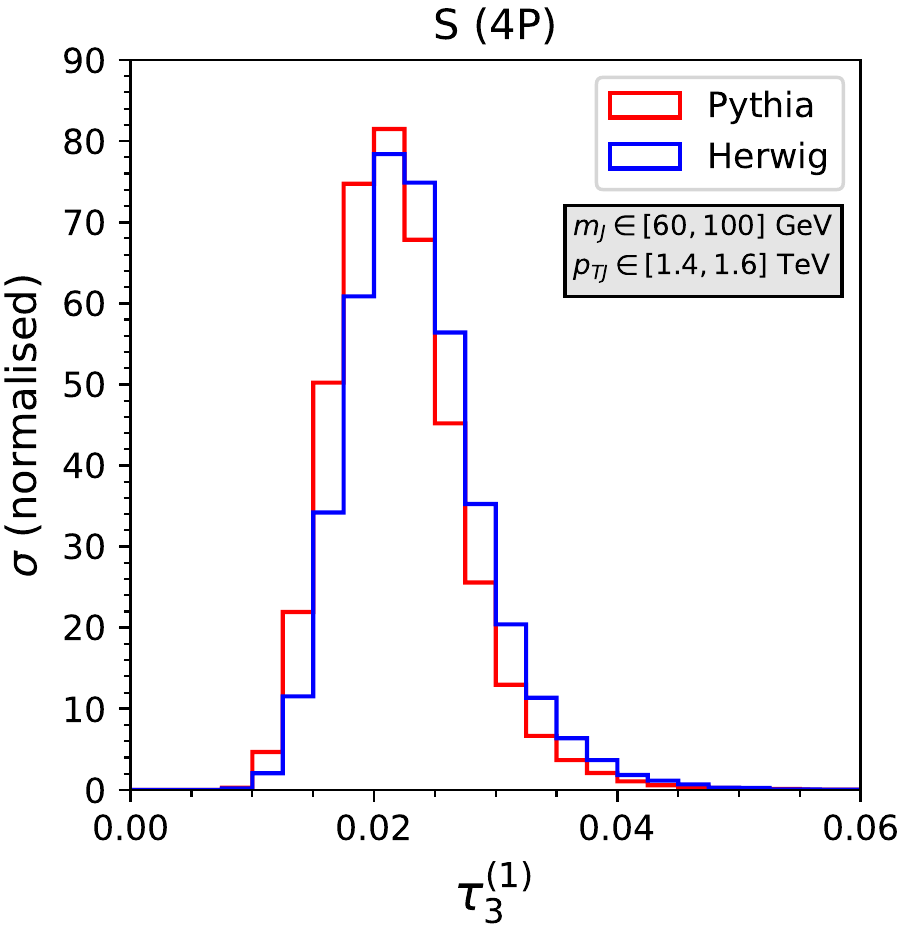} \\
\includegraphics[width=5.2cm,clip=]{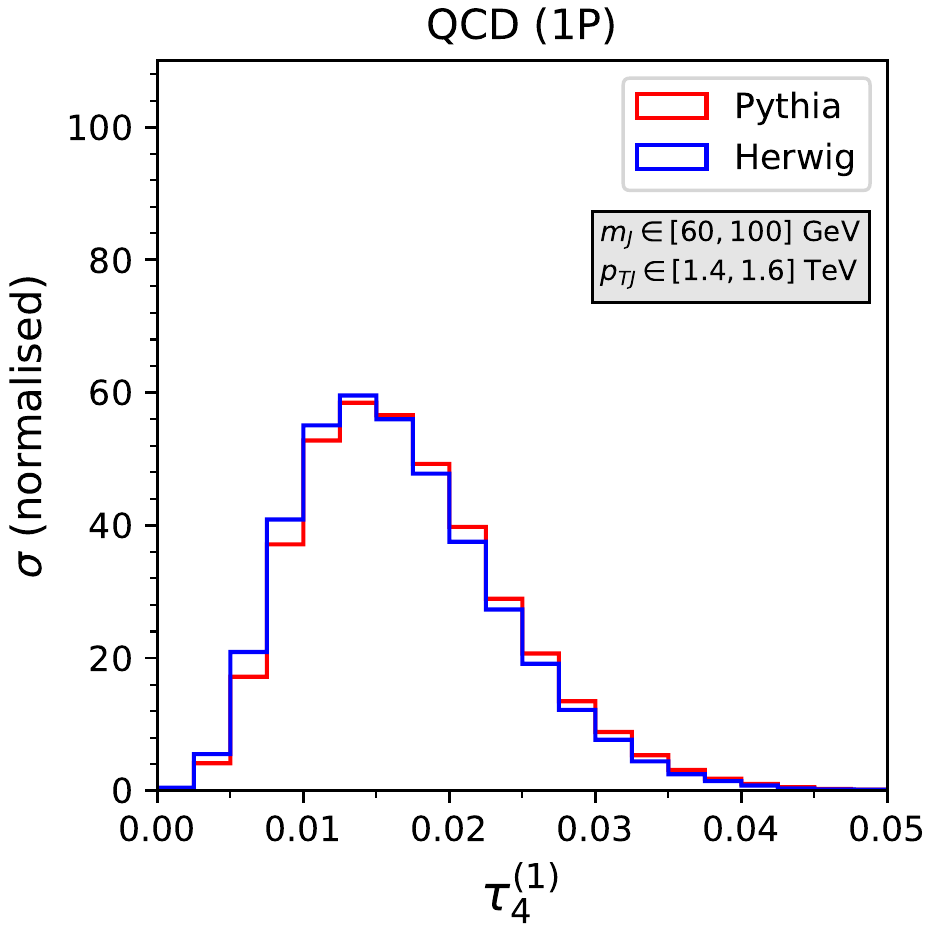} &
\includegraphics[width=5.2cm,clip=]{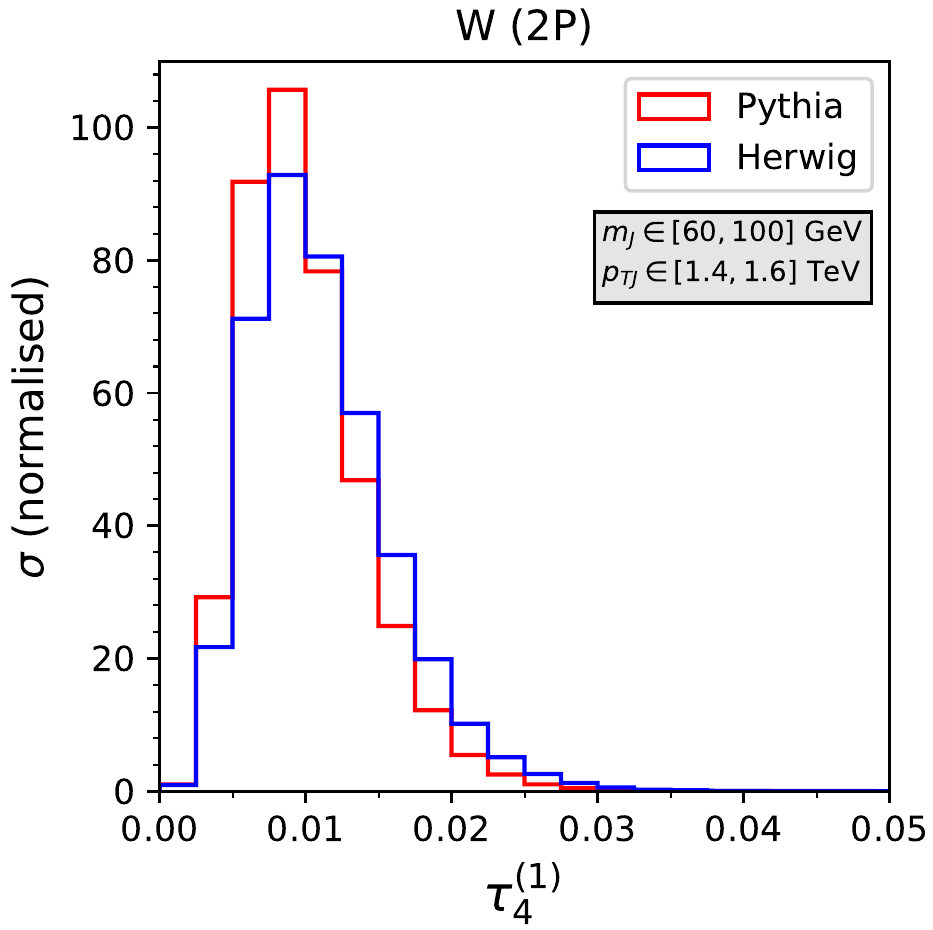} & 
\includegraphics[width=5.2cm,clip=]{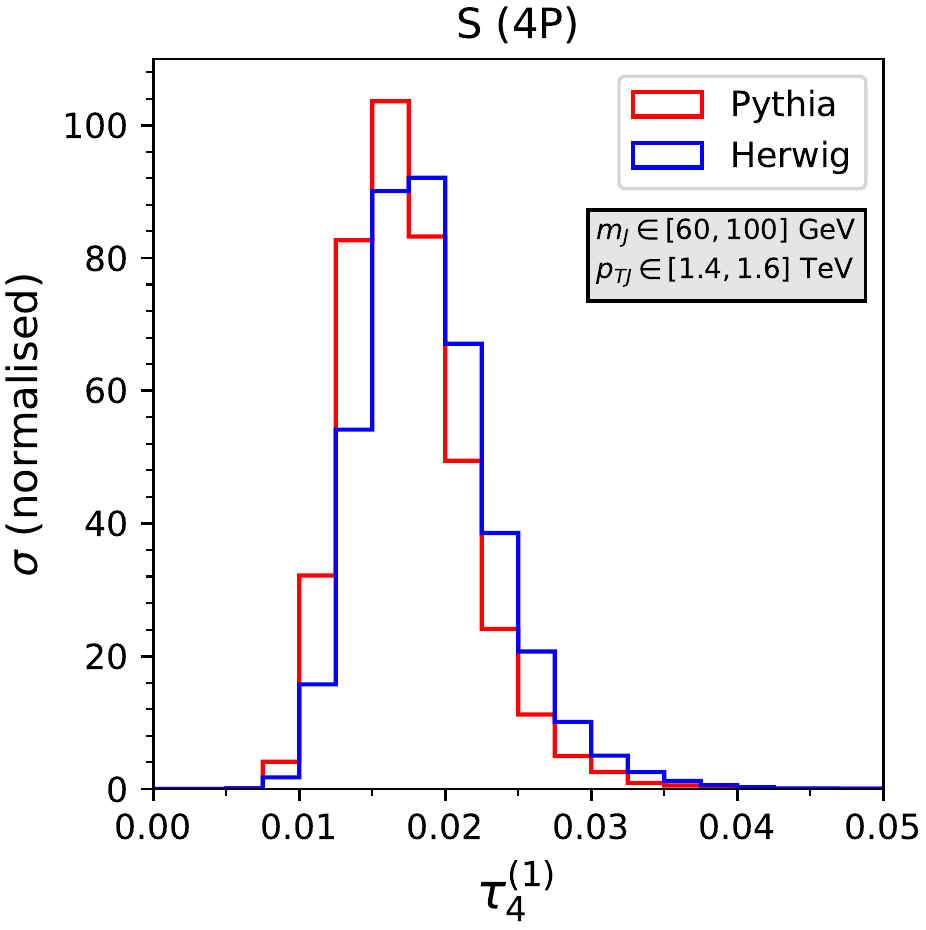} \\
\end{tabular}
\caption{Normalised distributions of $\tau_n^{(1)}$ with $n=1,2,3,4$, for QCD and multi-pronged jets with $\mj \in [60,100]$ GeV (see the text for details).}
\label{fig:tau80}
\end{center}
\end{figure*}

\begin{figure*}[p]
\begin{center}
\begin{tabular}{cccc}
\includegraphics[width=5.2cm,clip=]{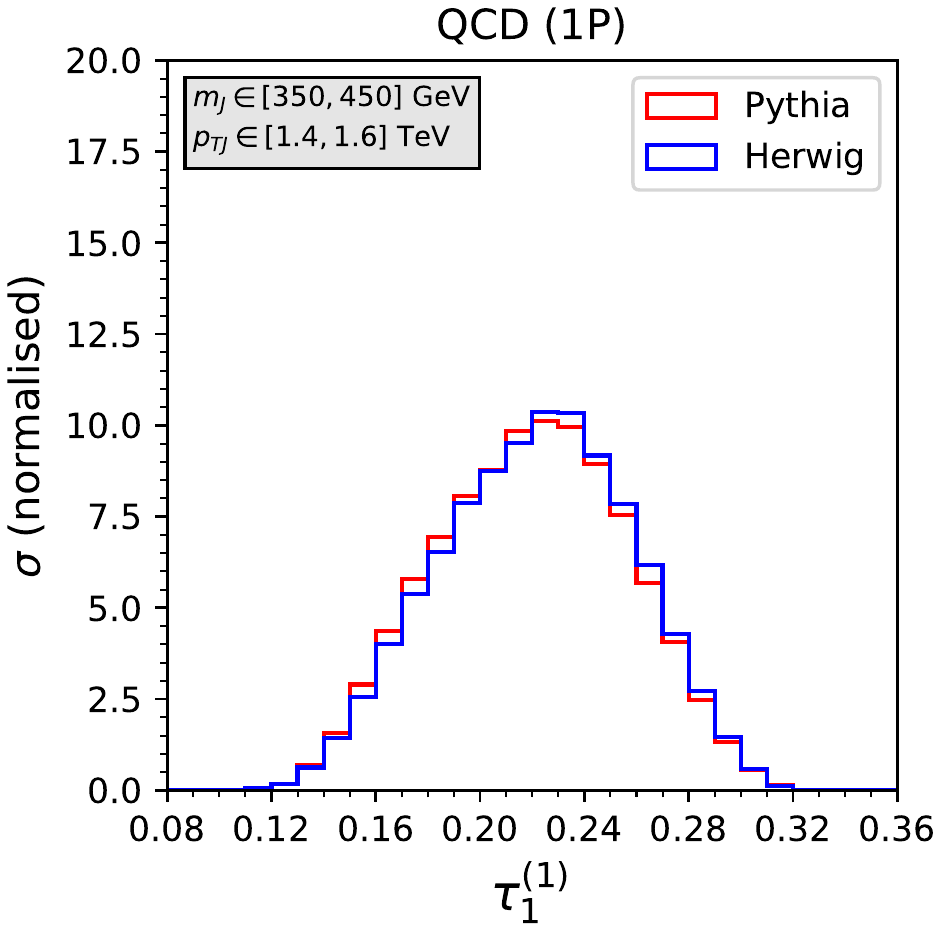} &
\includegraphics[width=5.2cm,clip=]{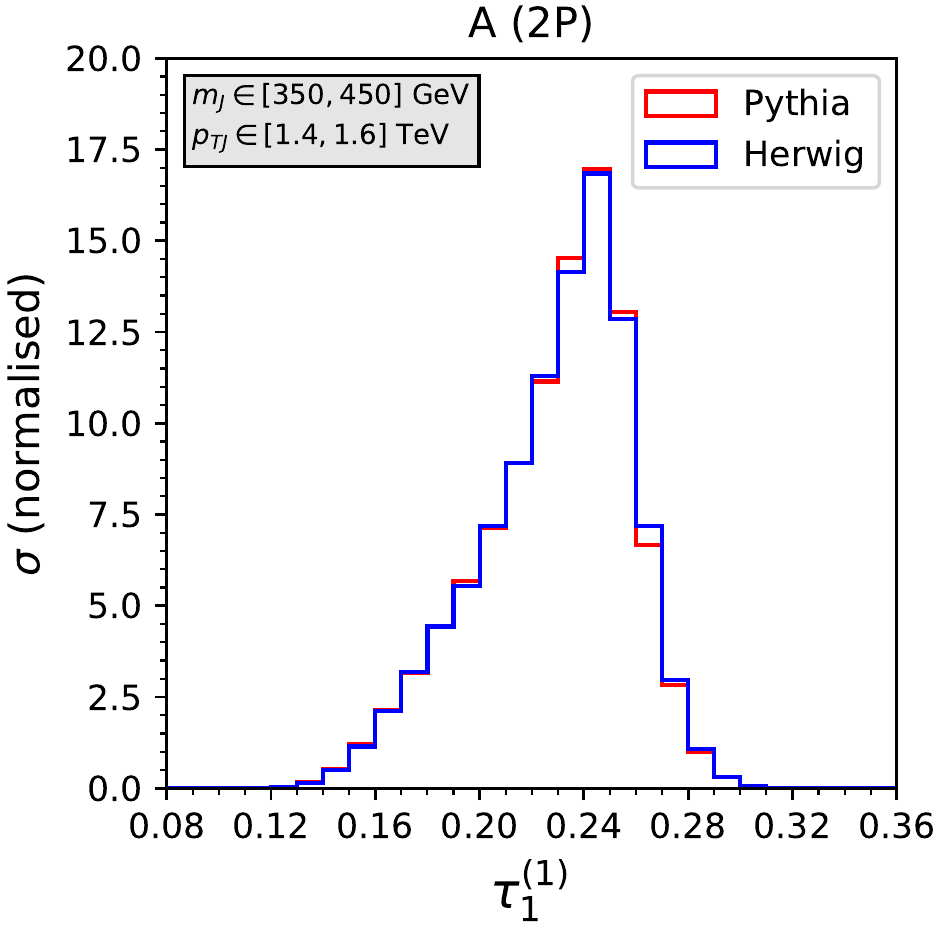} & 
\includegraphics[width=5.2cm,clip=]{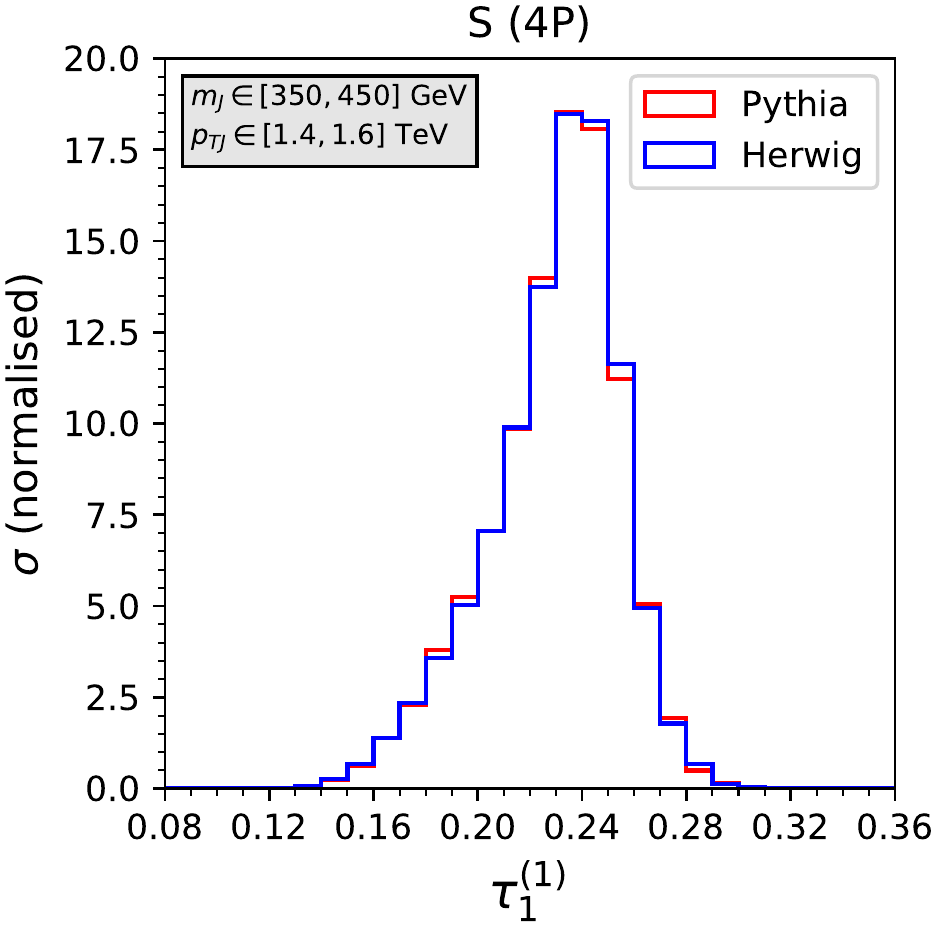} \\
\includegraphics[width=5.2cm,clip=]{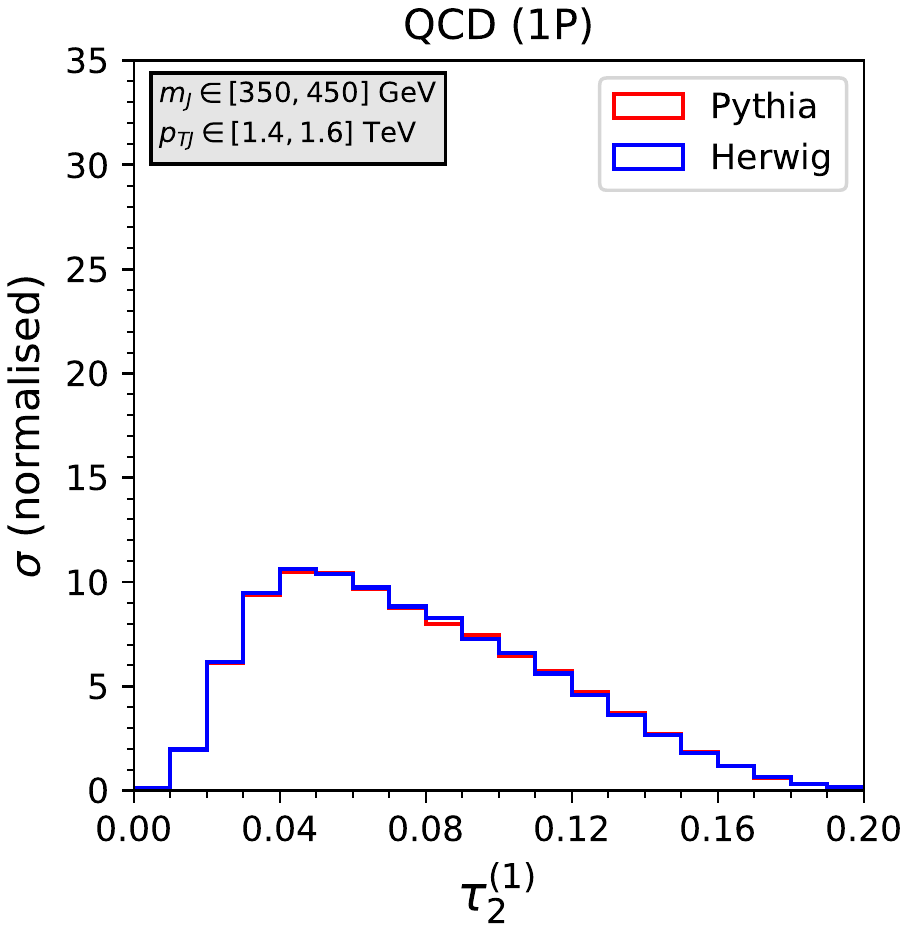} &
\includegraphics[width=5.2cm,clip=]{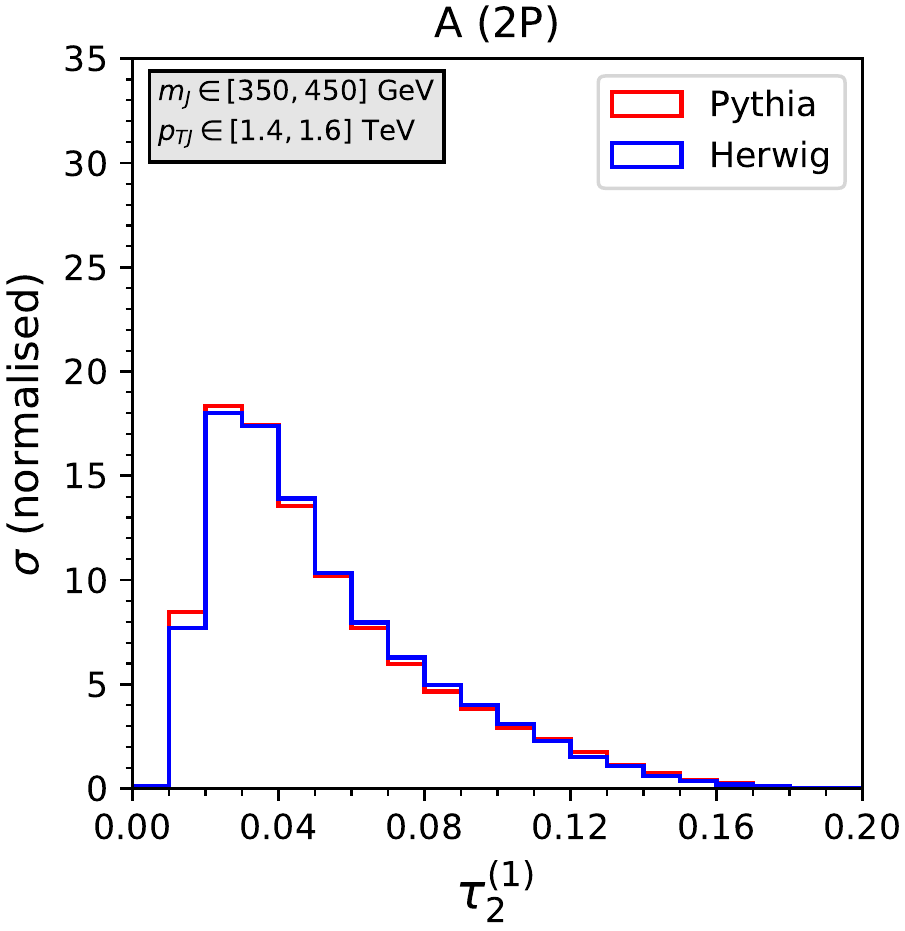} & 
\includegraphics[width=5.2cm,clip=]{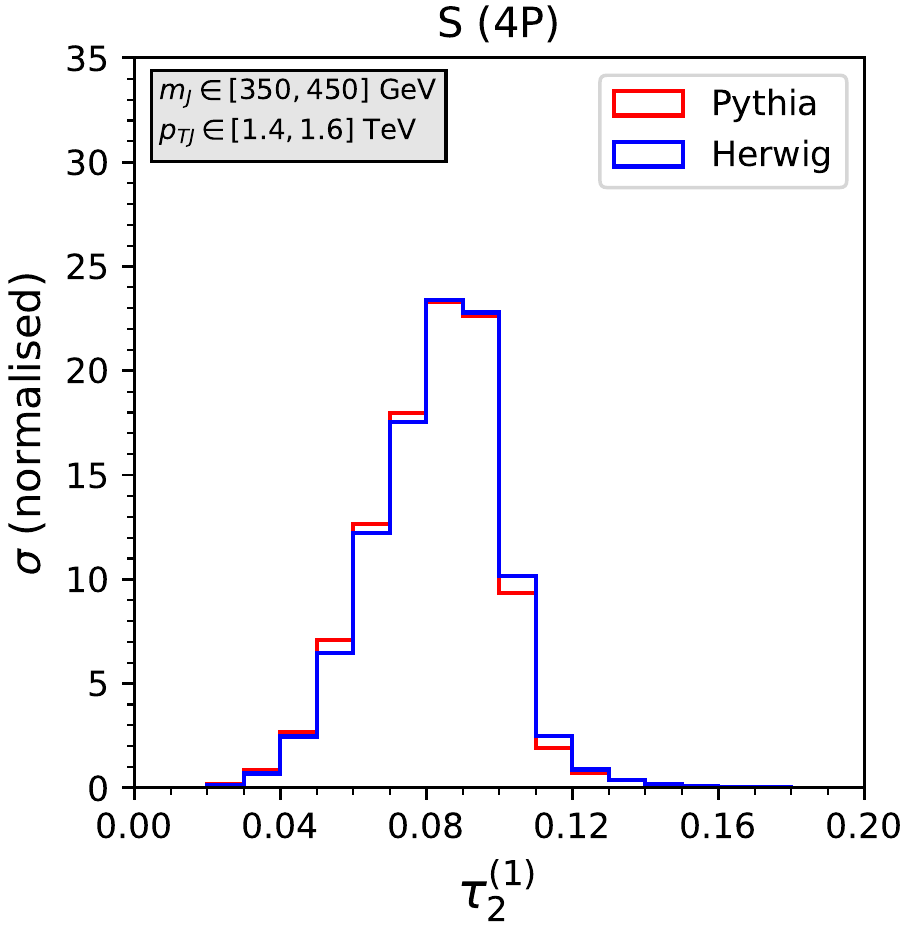} \\
\includegraphics[width=5.2cm,clip=]{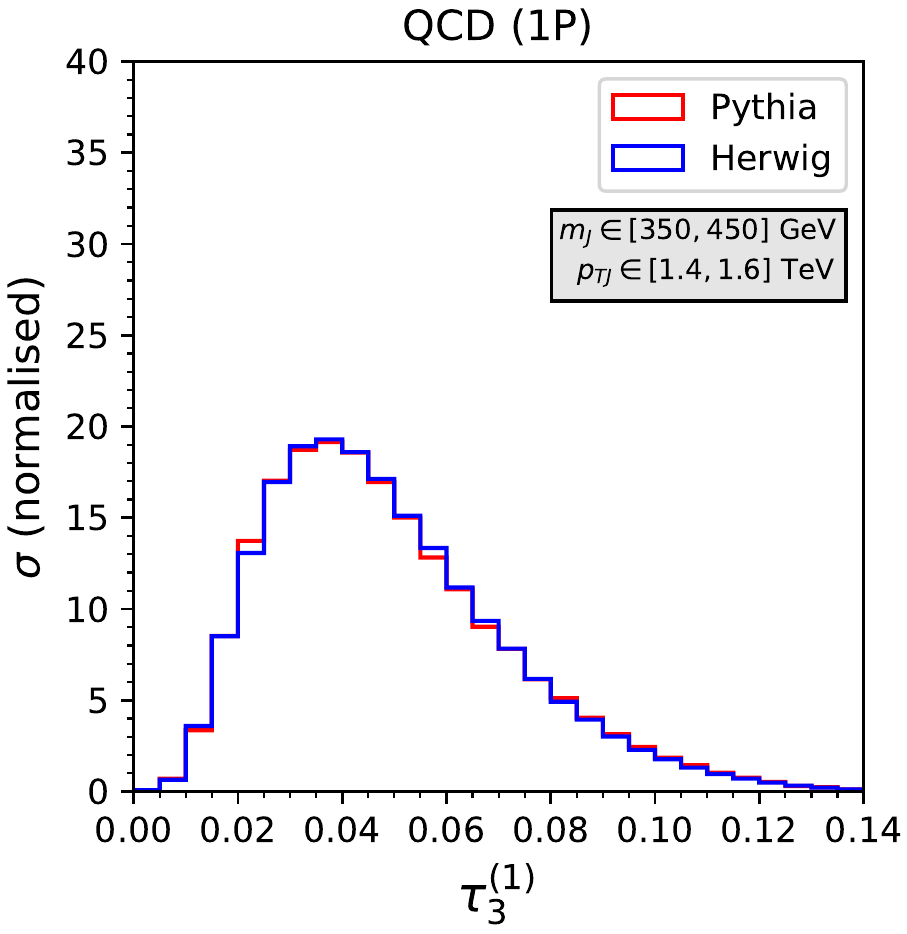} &
\includegraphics[width=5.2cm,clip=]{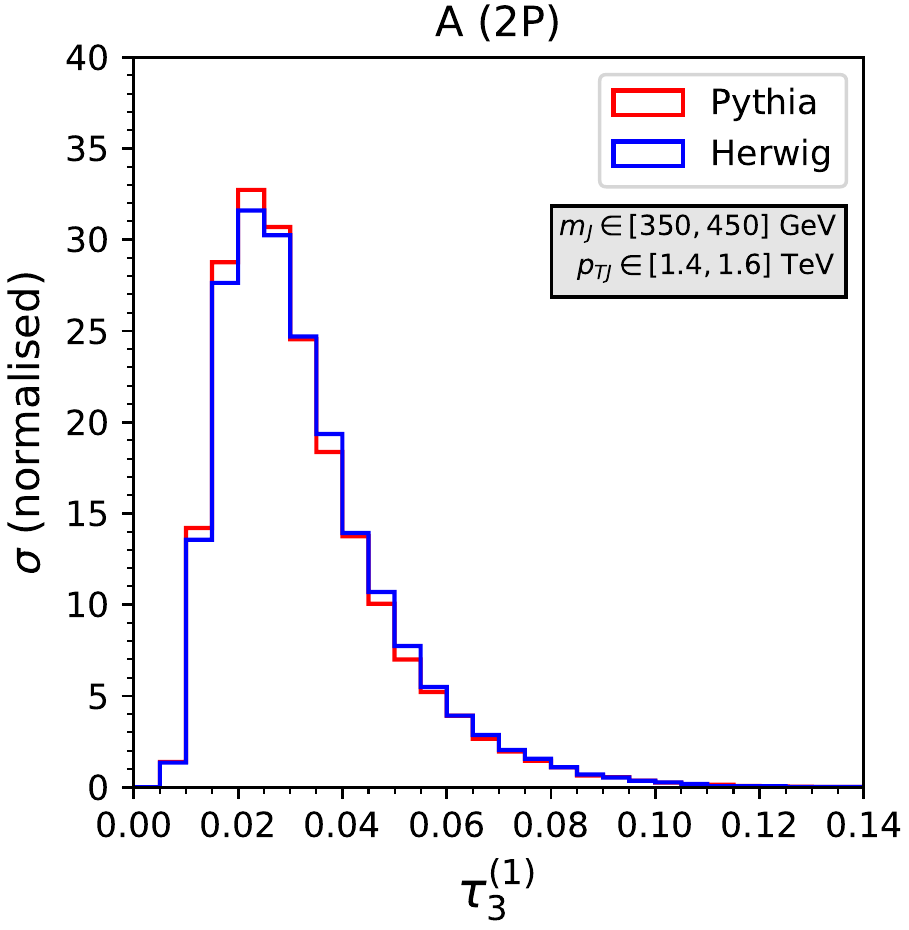} & 
\includegraphics[width=5.2cm,clip=]{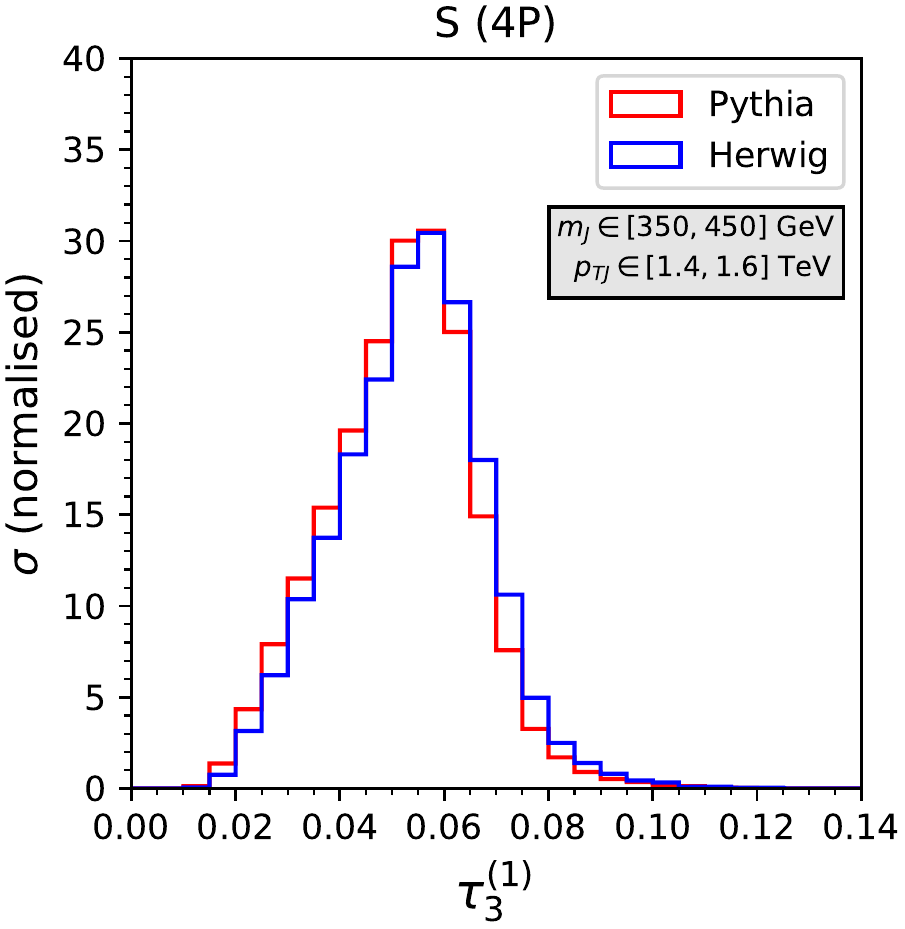} \\
\includegraphics[width=5.2cm,clip=]{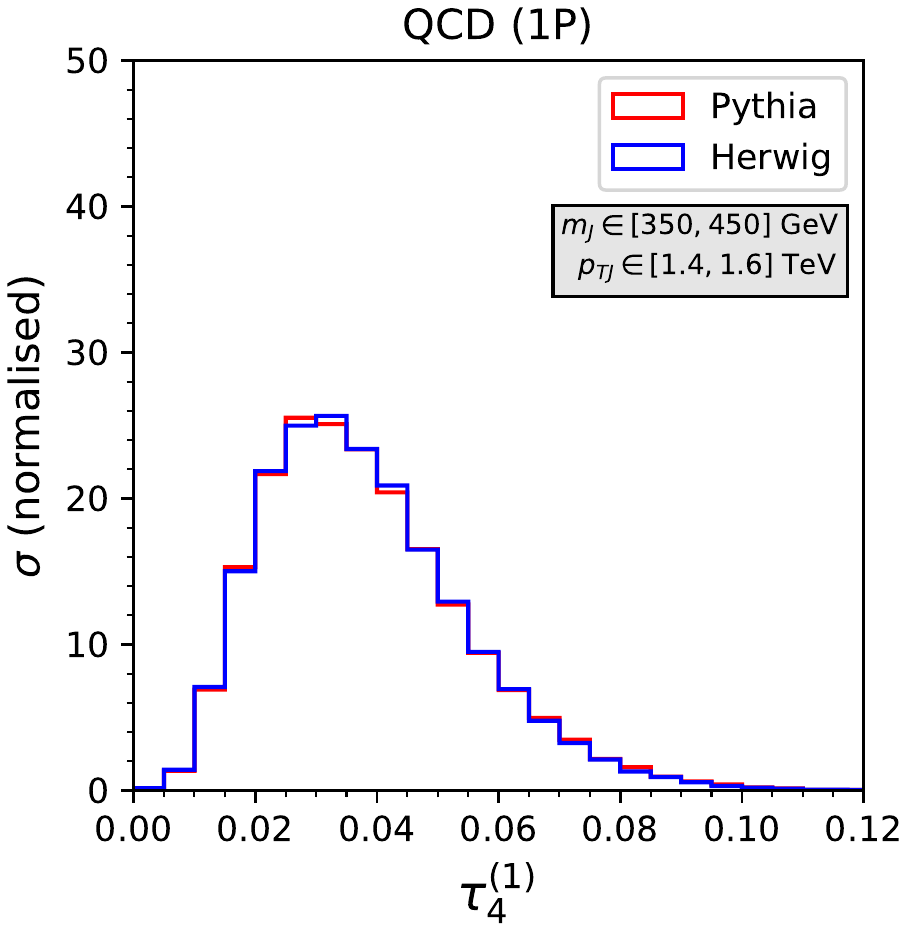} &
\includegraphics[width=5.2cm,clip=]{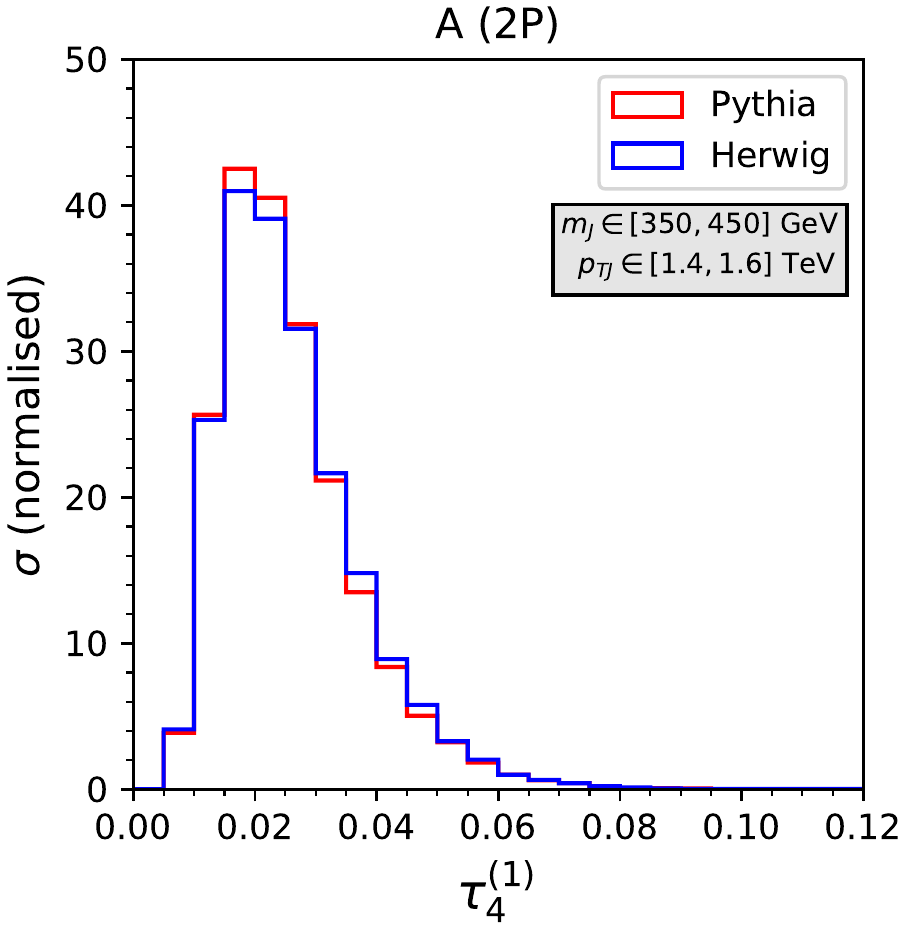} & 
\includegraphics[width=5.2cm,clip=]{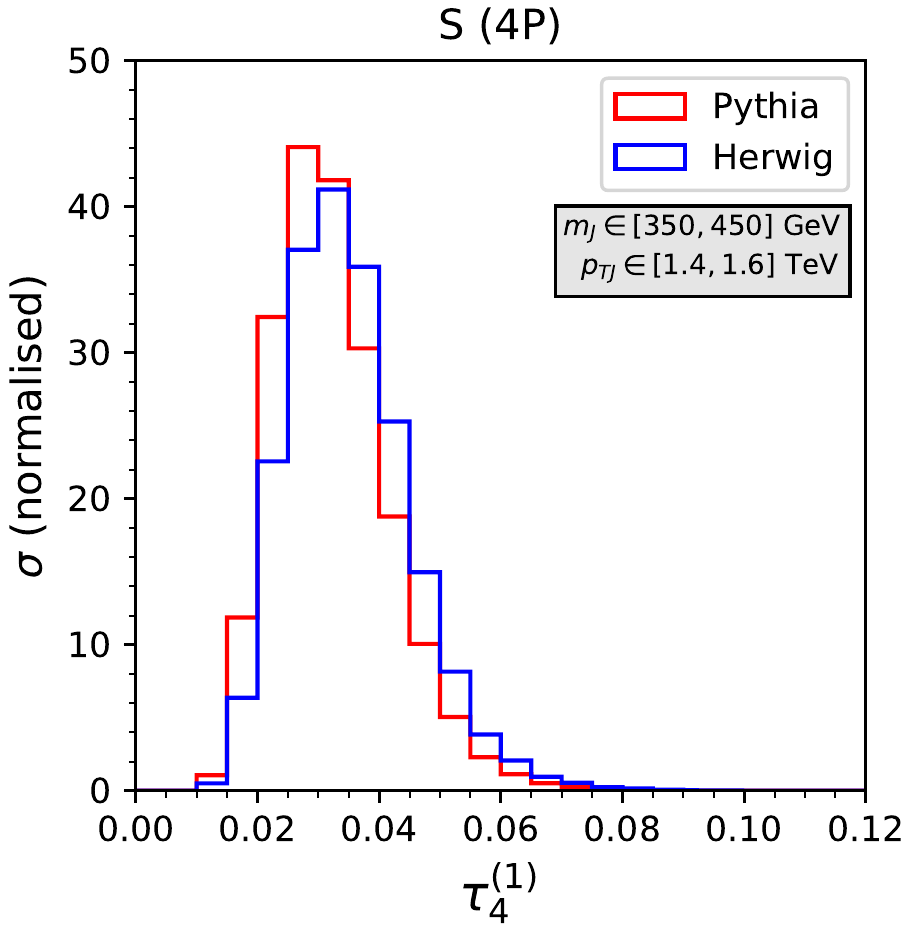} \\
\end{tabular}
\caption{Normalised distributions of $\tau_n^{(1)}$ with $n=1,2,3,4$, for QCD and multi-pronged jets with $\mj \in [350,450]$ GeV (see the text for details).}
\label{fig:tau400}
\end{center}
\end{figure*}

For the comparison we consider QCD and multi-pronged jets with $\ptj \in [1.4,1.6]$ TeV and two ranges for the jet masses: (a) $\mj \in [60,100]$ GeV; (b) $\mj \in [350,450]$ GeV. The multi-pronged jets are generated from the decay of a 3.3 TeV $Z'$:
\begin{itemize}
\item For $\mj \in [60,100]$ GeV we use $Z' \to WW$, $W \to q \bar q$ (two-pronged, 2P) and $Z' \to SS$, $S \to AA \to 4b$ with $M_S = 80$ GeV, $M_A = 30$ GeV (four-pronged, 4P)
\item For $\mj \in [350,450]$ GeV we use $Z' \to AA$, $A \to b \bar b$ with $M_A = 400$ GeV (2P) and 
$Z' \to SS$, $S \to AA \to 4b$ with $M_S = 400$ GeV, $M_A = 80$ GeV (4P).
\end{itemize}
We present in Fig.~\ref{fig:tau80} the normalised distributions of $\tau_n^{(1)}$ with $n=1,2,3,4$, for QCD and multi-pronged jets with $\mj \in [60,100]$ GeV. The results for $\mj \in [350,450]$ GeV are displayed in Fig.~\ref{fig:tau400}.

The distributions for $n=1$ (top rows) are quite the same when using {\scshape Pythia} (red) or {\scshape Herwig} (blue). For $n = 2$ there is some difference, which increases with $n$ for 2P and 4P jets. 
Furthermore, for 2P and 4P jets the level of (dis)agreement between {\scshape Pythia} and {\scshape Herwig} distributions is alike, with the exception of $\tau_4^{(1)}$ for $\mj \in [350,450]$ GeV.
Because the discrimination between 2P jets and the QCD background is less sensitive to higher-order $\tau_n^{(i)}$, one then expects that the differences in tagger performance found for 2P jets will be milder. This is confirmed by the results presented in the next section. In addition, we remark that the differences are more significant for
lighter boosted particles, i.e. $\mj \in [60,100]$ GeV. 

We also observe that in all cases the {\scshape Pythia} and {\scshape Herwig} distributions for QCD jets are very similar. On the other hand, {\scshape Herwig} distributions for $\tau_n^{(1)}$, $n \geq 2$ are slightly shifted towards larger values for 2P and 4P jets. This fact indicates that in 2P and 4P jets the subjets are less resolved, and is 
in agreement with Ref.~\cite{Barnard:2016qma}, which shows that {\scshape Herwig} with angular-ordered shower produces more diffuse radiation and less resolved subjets for $W$ jets.
This pattern also shows that the discrepancy is mainly due to the hadronisation scheme, rather than parton showering. The results in Ref.~\cite{Barnard:2016qma} confirm this point: the differences that arise when using the same hadronisation scheme but different showering (e.g.  {\scshape Herwig} with either angular-ordered or dipole shower;  {\scshape Pythia} with either dipole or antenna showers) are quite small. 

\begin{figure*}[t!]
\begin{center}
\begin{tabular}{cc}
\includegraphics[width=8cm,clip=]{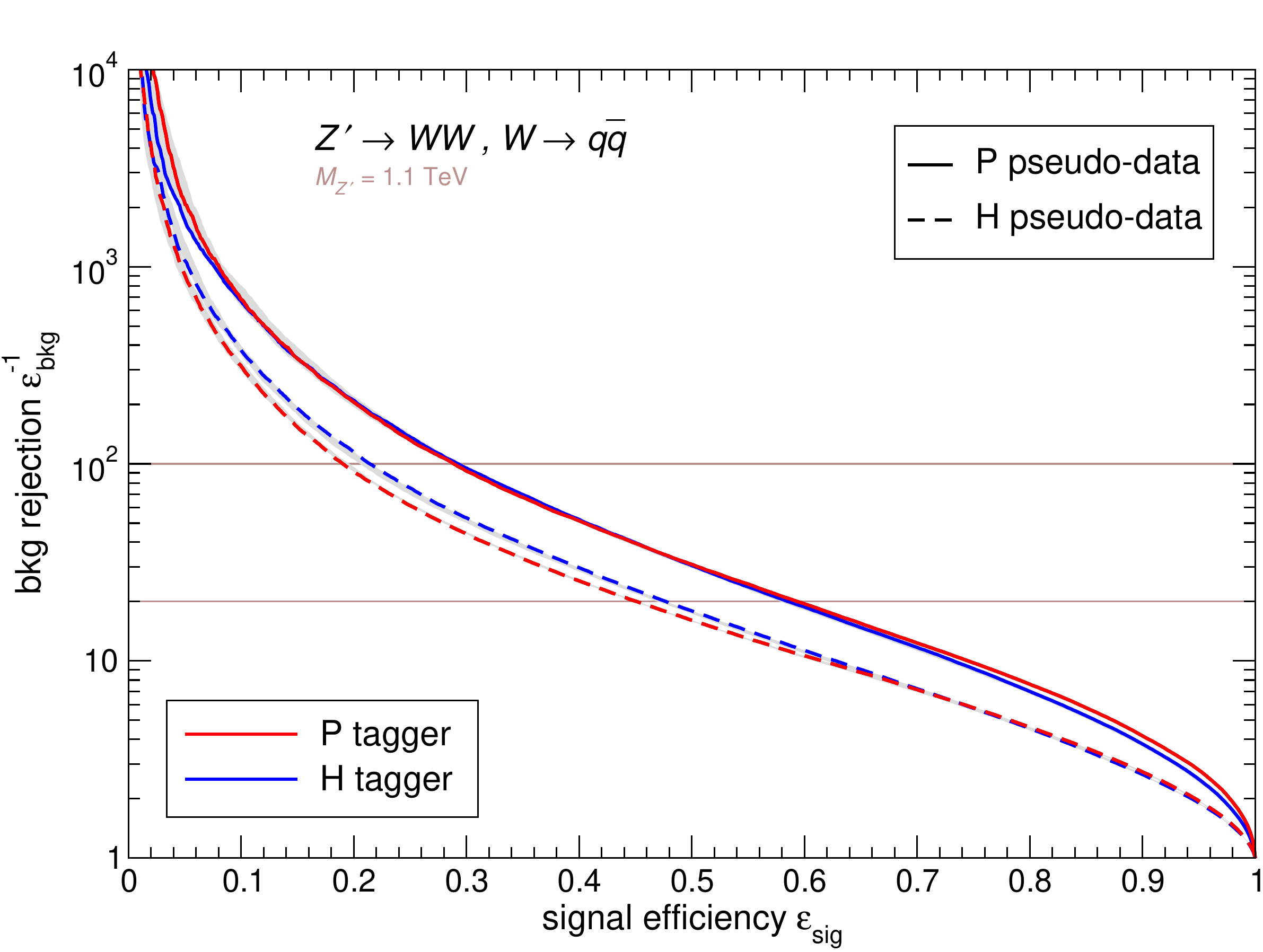} 
& \includegraphics[width=8cm,clip=]{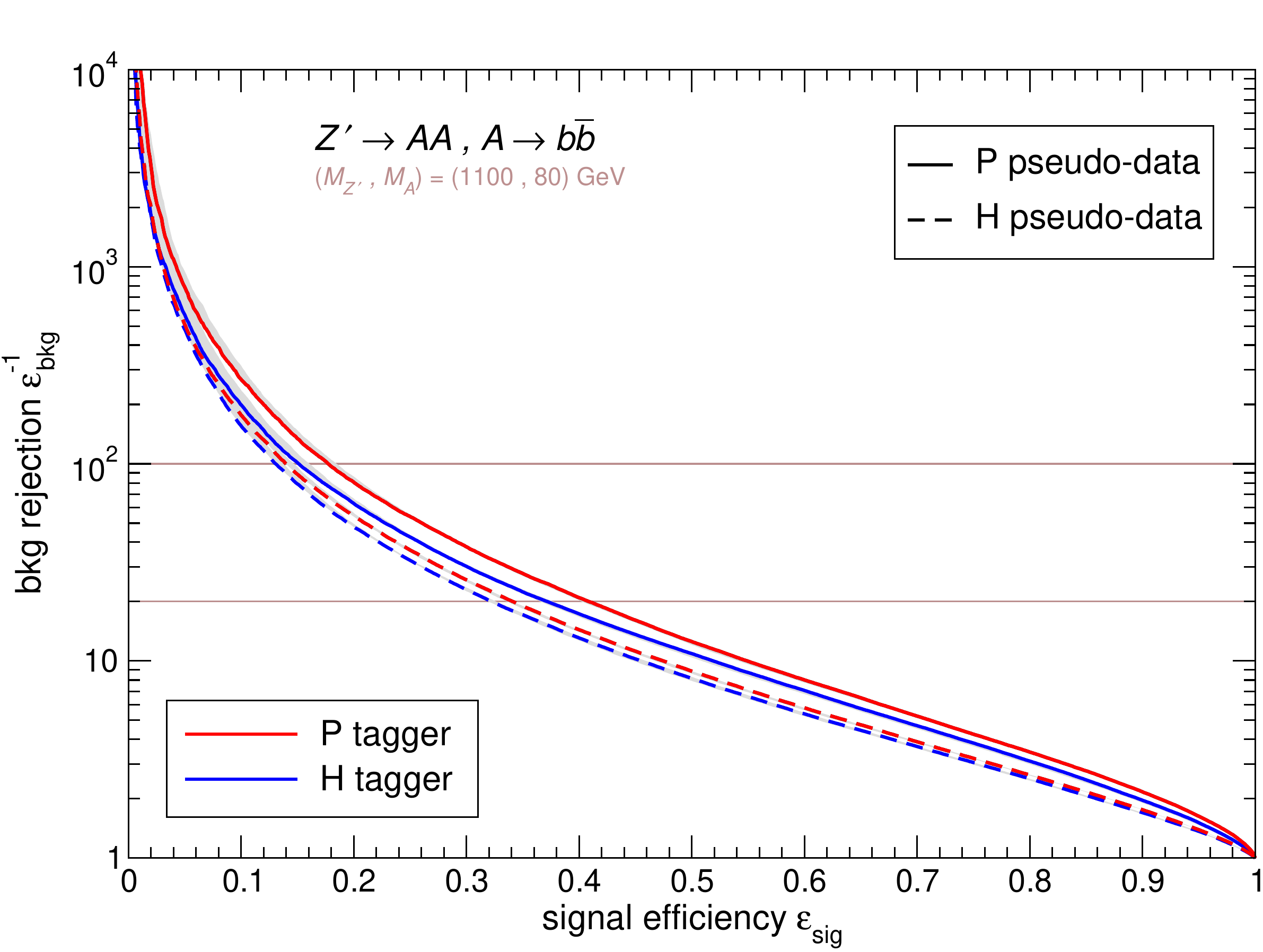} \\
\includegraphics[width=8cm,clip=]{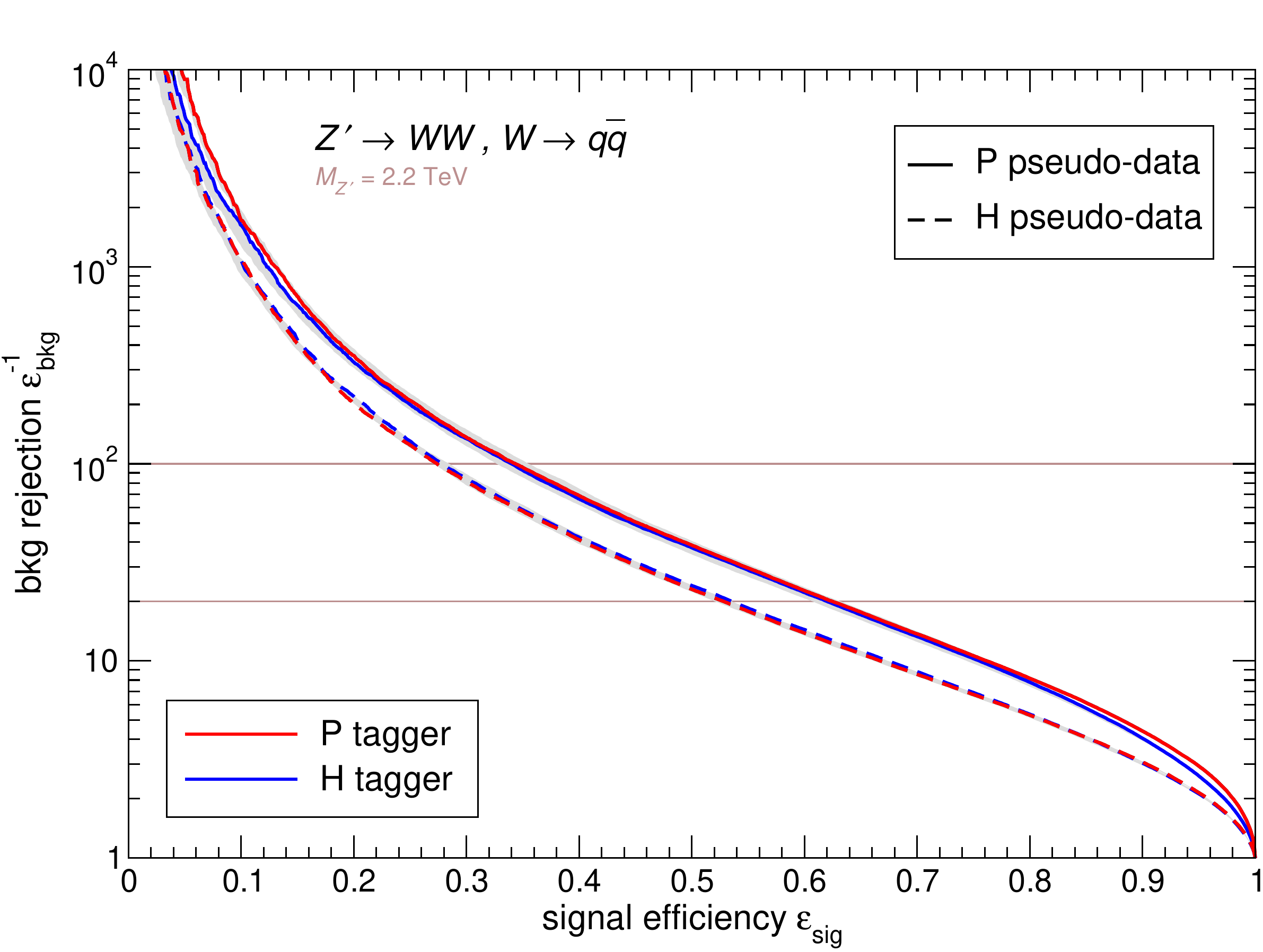}
& \includegraphics[width=8cm,clip=]{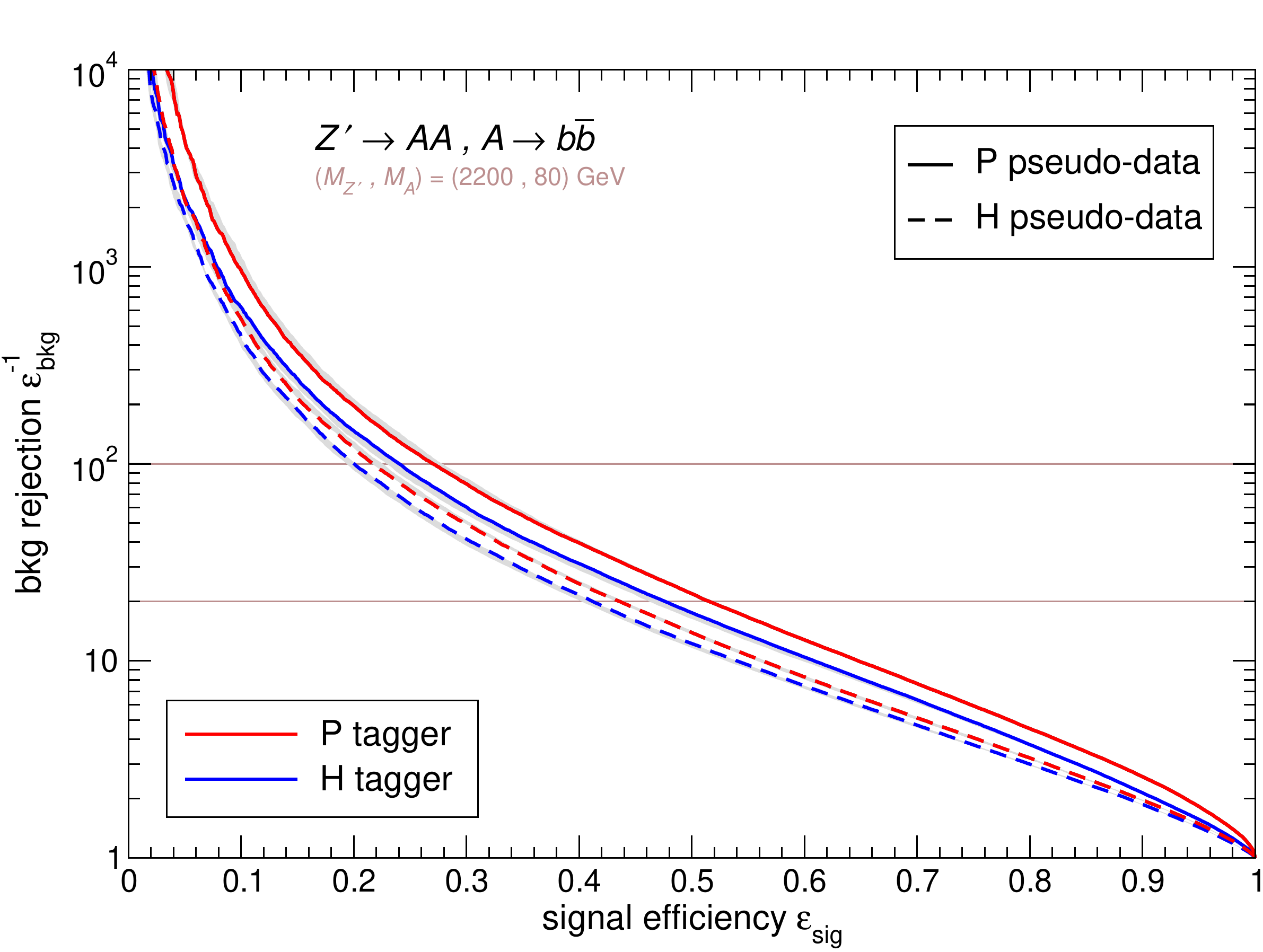} \\
\includegraphics[width=8cm,clip=]{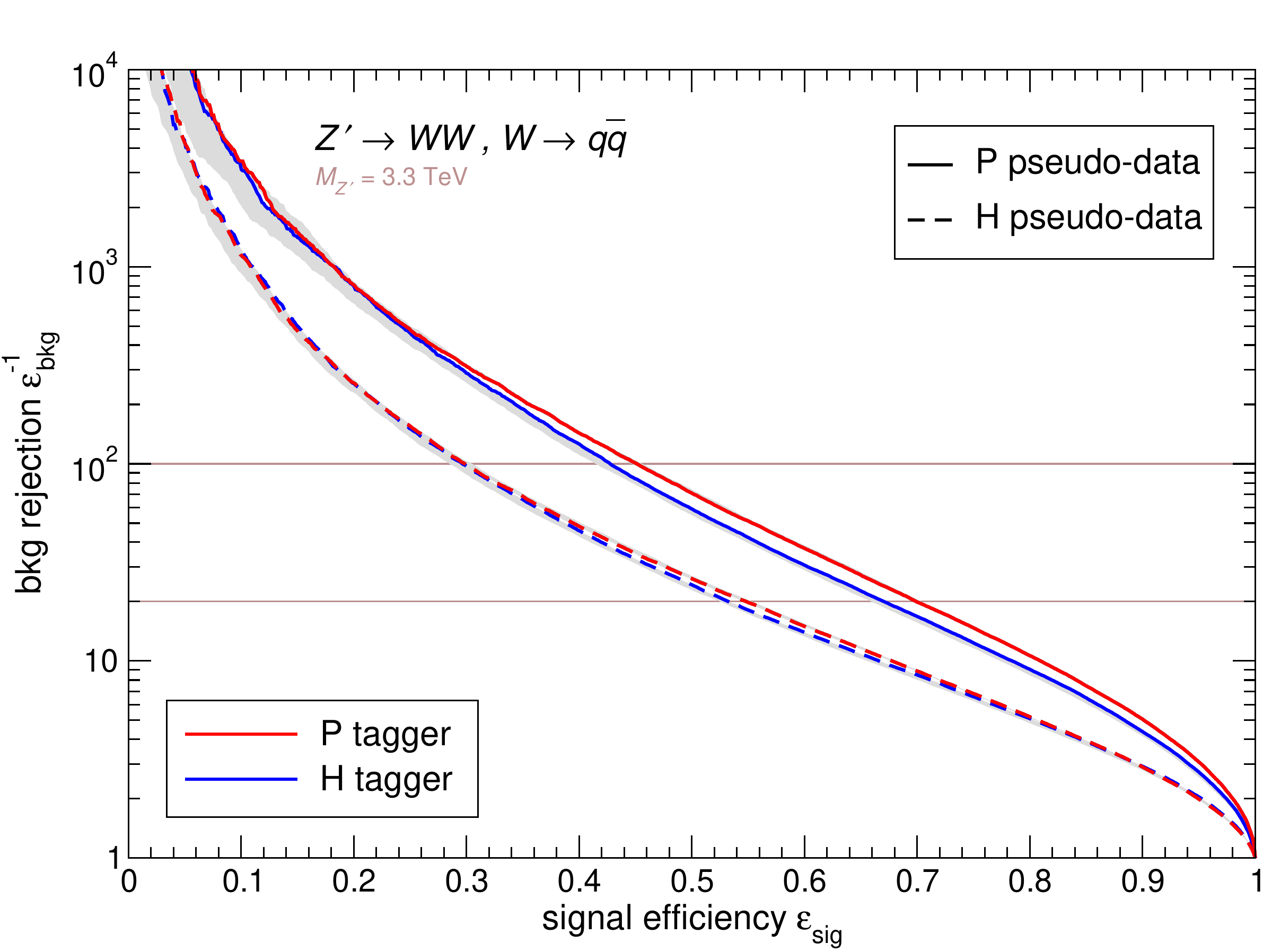}
& \includegraphics[width=8cm,clip=]{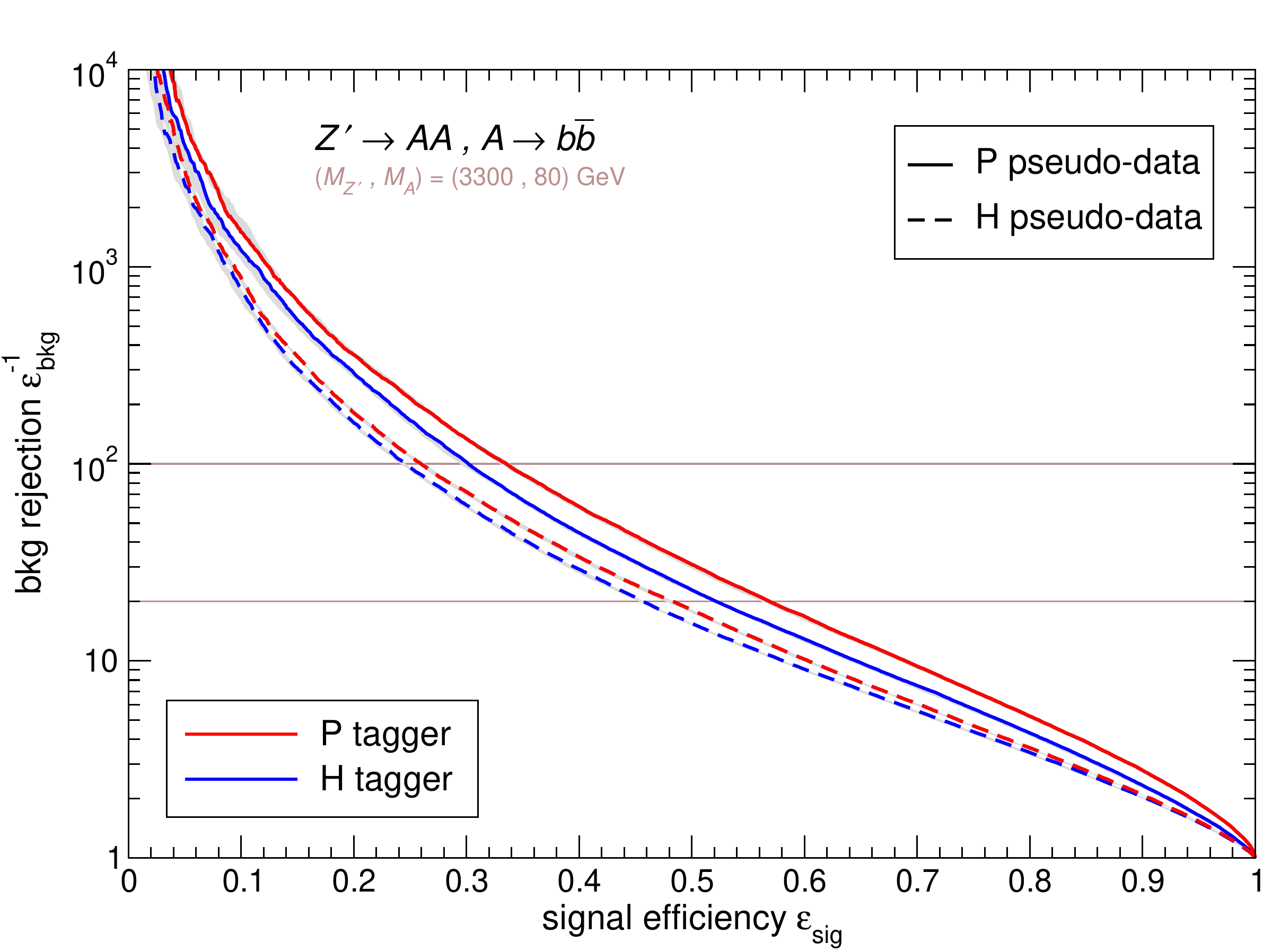} 
\end{tabular}
\caption{ROC curves 2P jets with $\mj \sim 80$ GeV.}
\label{fig:ROC-2P80}
\end{center}
\end{figure*}

\begin{figure}[t]
\begin{center}
\begin{tabular}{c}
\includegraphics[width=8cm,clip=]{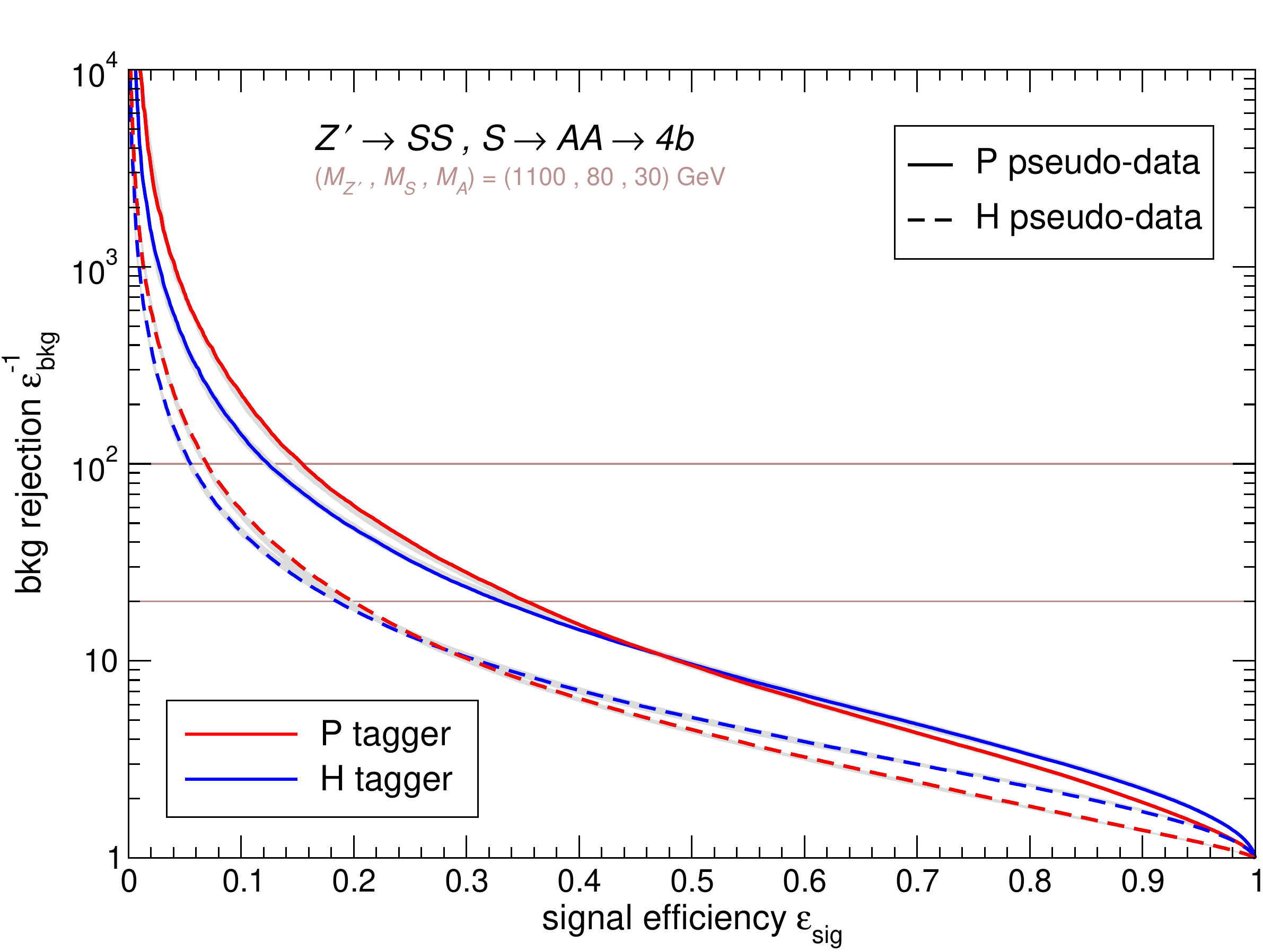} \\
\includegraphics[width=8cm,clip=]{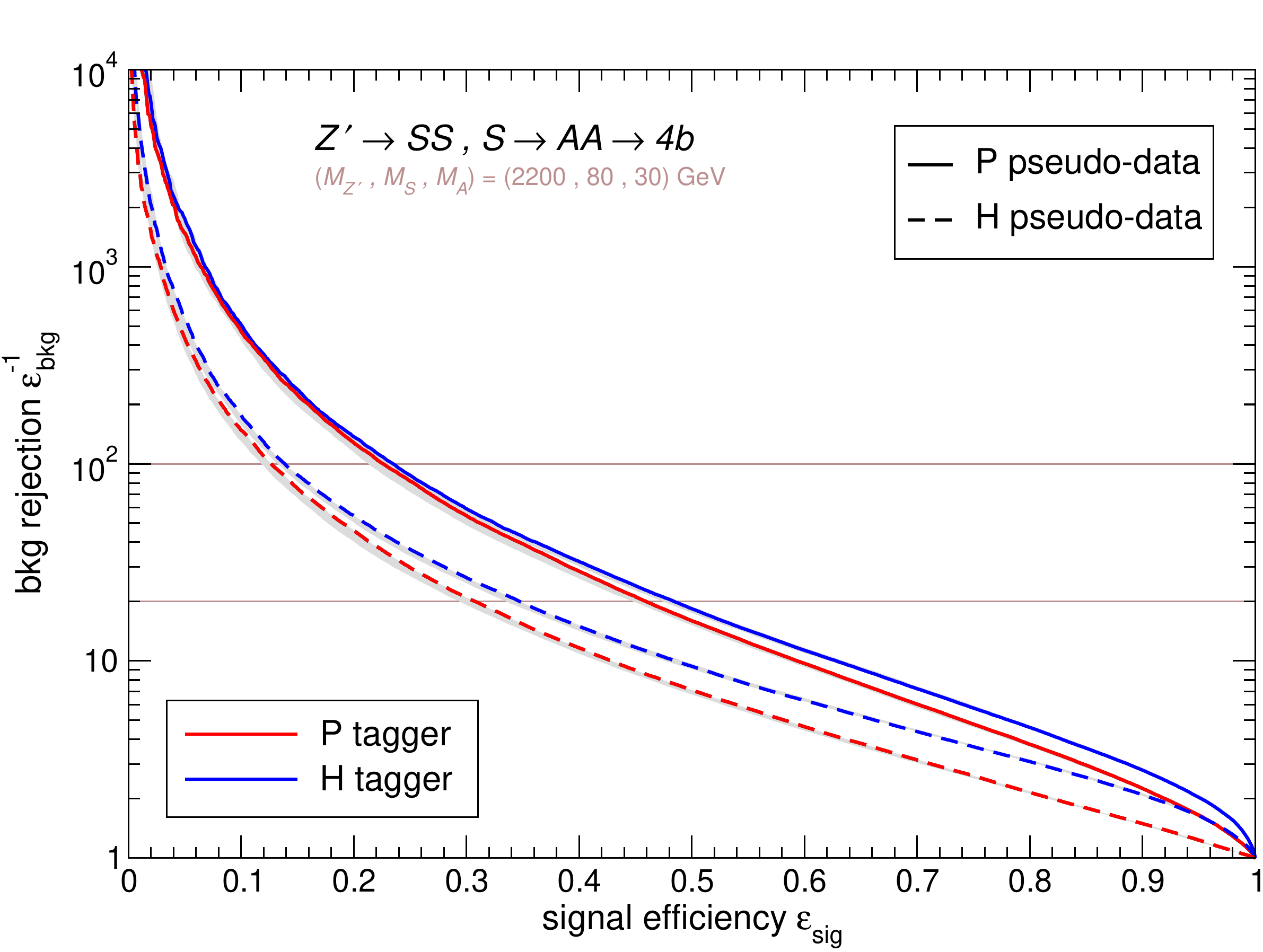} \\
\includegraphics[width=8cm,clip=]{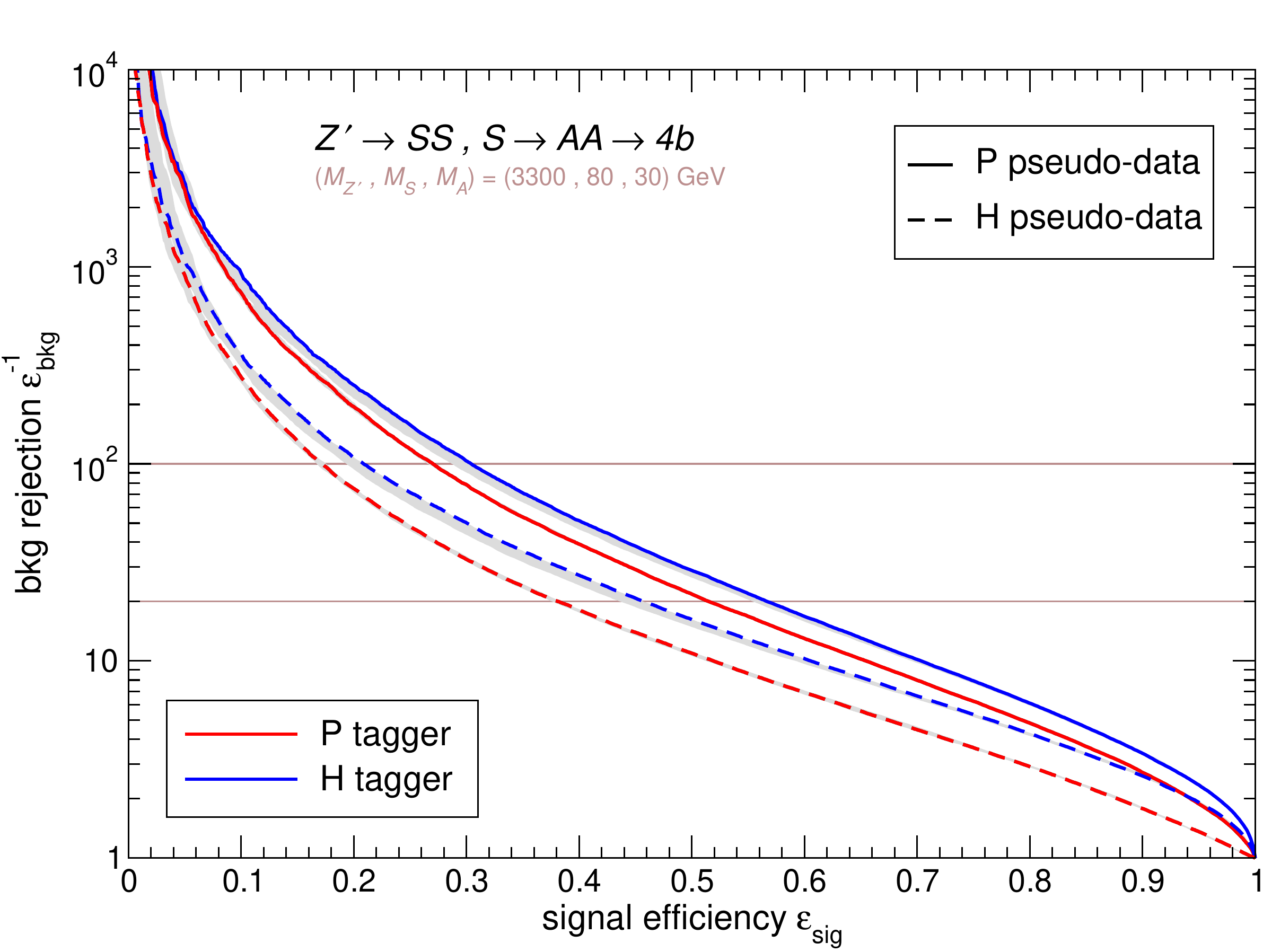} 
\end{tabular}
\caption{ROC curves for 4P jets with $\mj \sim 80$ GeV.}
\label{fig:ROC-4P80}
\end{center}
\end{figure} 

\section{Tagging performance}
\label{sec:4}

In this section we test the two taggers trained either with {\scshape Pythia} (P) or {\scshape Herwig} (H) on pseudo-data, generated with either of these Monte Carlo simulations. This allows to disentangle two important aspects that are independent:
\begin{itemize}
 \item[(a)] the modeling dependence, that is, the different performance of the two taggers when applied to the same pseudo-data;
 \item[(b)] the dependence of the performance on pseudo-data itself, that is, applying the same tagger to different pseudo-data.
 \end{itemize}
We consider the five signal processes mentioned in section~\ref{sec:2} with different masses, totaling 18 benchmarks:
\begin{itemize}
\item[(1)] $Z' \to WW$, with $M_{Z'} = 1.1$, 2.2, 3.3 TeV.
\item[(2)] $Z' \to AA$, with $(M_{Z'},M_A) =$ (1100, 80), (2200, 80), (3300, 80), (3300, 400) GeV.
\item[(3)] $Z' \to t \bar t$ with $M_{Z'} =$ 2.2, 3.3 TeV.
\item[(4)] $Z' \to SS$, $S \to WW$ with  $(M_{Z'},M_S) =$ (2200, 200),  (3300, 200), (3300, 400) GeV. 
\item[(5)] $Z' \to SS$, $S \to AA$ with $(M_{Z'},M_S,M_A) =$ (1100, 80, 30), (2200, 80, 30), (3300, 80, 30), (2200, 200, 80), (3300, 200, 80), (3300, 400, 80) GeV. 
\end{itemize}
For the benchmarks with $M_{Z'} = 1.1$, 2.2, 3.3 TeV we select jets with transverse momentum within the respective intervals $\ptj \in [0.4,0.6]$, $[0.9,1.1]$, $[1.4,1.6]$ TeV. For the benchmarks with jet mass $\mj \sim 80$, 175, 200, 400 GeV we select jets with mass in the respective intervals $\mj \in [60,100]$, $[150,200]$, $[160,240]$, $[350,450]$ GeV.

We first present in Fig.~\ref{fig:ROC-2P80} the receiver operating characteristic (ROC) curves for light 2P jets: $Z' \to WW$ and $Z' \to AA$ with $M_A = 80$ GeV, with $M_{Z'} = 1.1$, 2.2 and 3.3 TeV. Figure~\ref{fig:ROC-4P80} shows results for 4P jets of the same mass, using $Z' \to SS$ with $M_S = 80$ GeV, $M_A = 30$ GeV, and the same $Z'$ masses. The gray bands around the curves represent the variation of $(\varepsilon_\text{sig},\varepsilon_\text{bkg}^{-1})$ among the five trainings of the NN. Except at the upper left side, the bands are barely visible and their width is comparable to the thickness of the curves. Showing the variation of $(\varepsilon_\text{sig},\varepsilon_\text{bkg}^{-1})$ among trainings can be used to test whether the difference between P and H taggers is a statistical artifact. We also include horizontal lines at $\varepsilon_\text{bkg}^{-1} = 20$, 100 to guide the eye to estimate the variation in $\varepsilon_\text{sig}$ between the different curves.

\begin{figure}[t]
\begin{center}
\begin{tabular}{c}
\includegraphics[width=8cm,clip=]{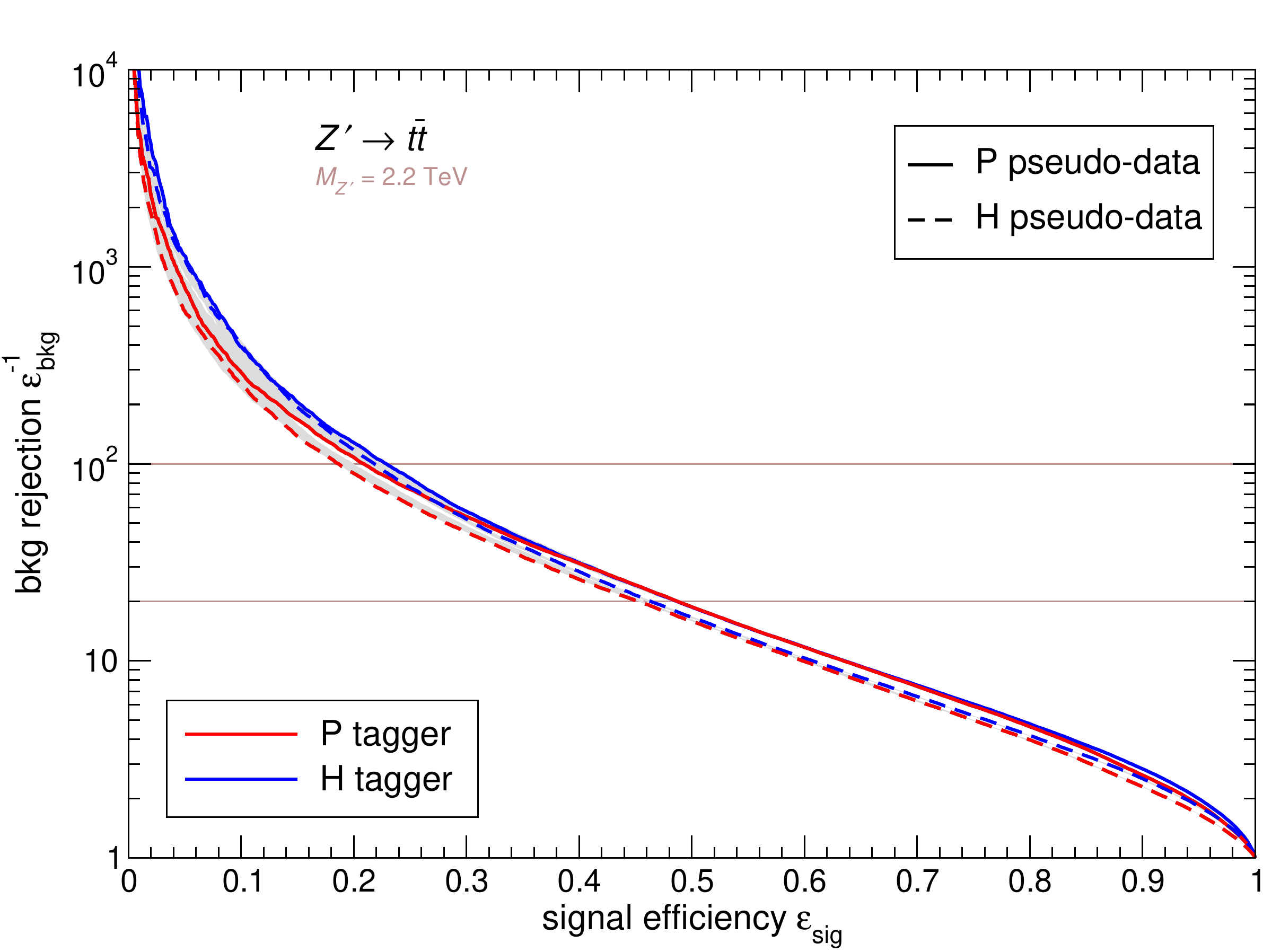} \\
\includegraphics[width=8cm,clip=]{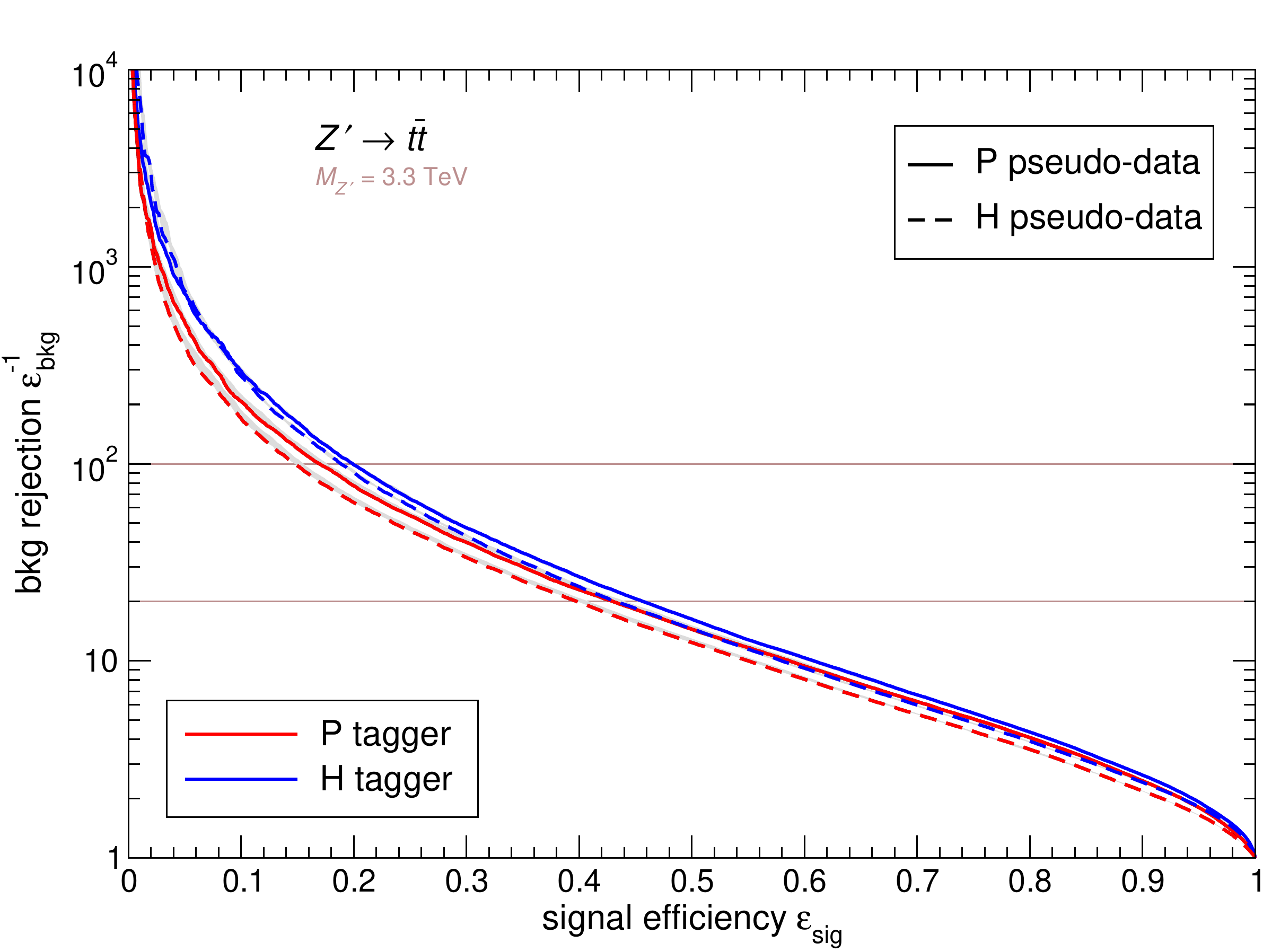} 
\end{tabular}
\caption{ROC curves for 3P top jets.}
\label{fig:ROC-3P}
\end{center}
\end{figure} 

\begin{figure*}[t!]
\begin{center}
\begin{tabular}{cc}
\includegraphics[width=8cm,clip=]{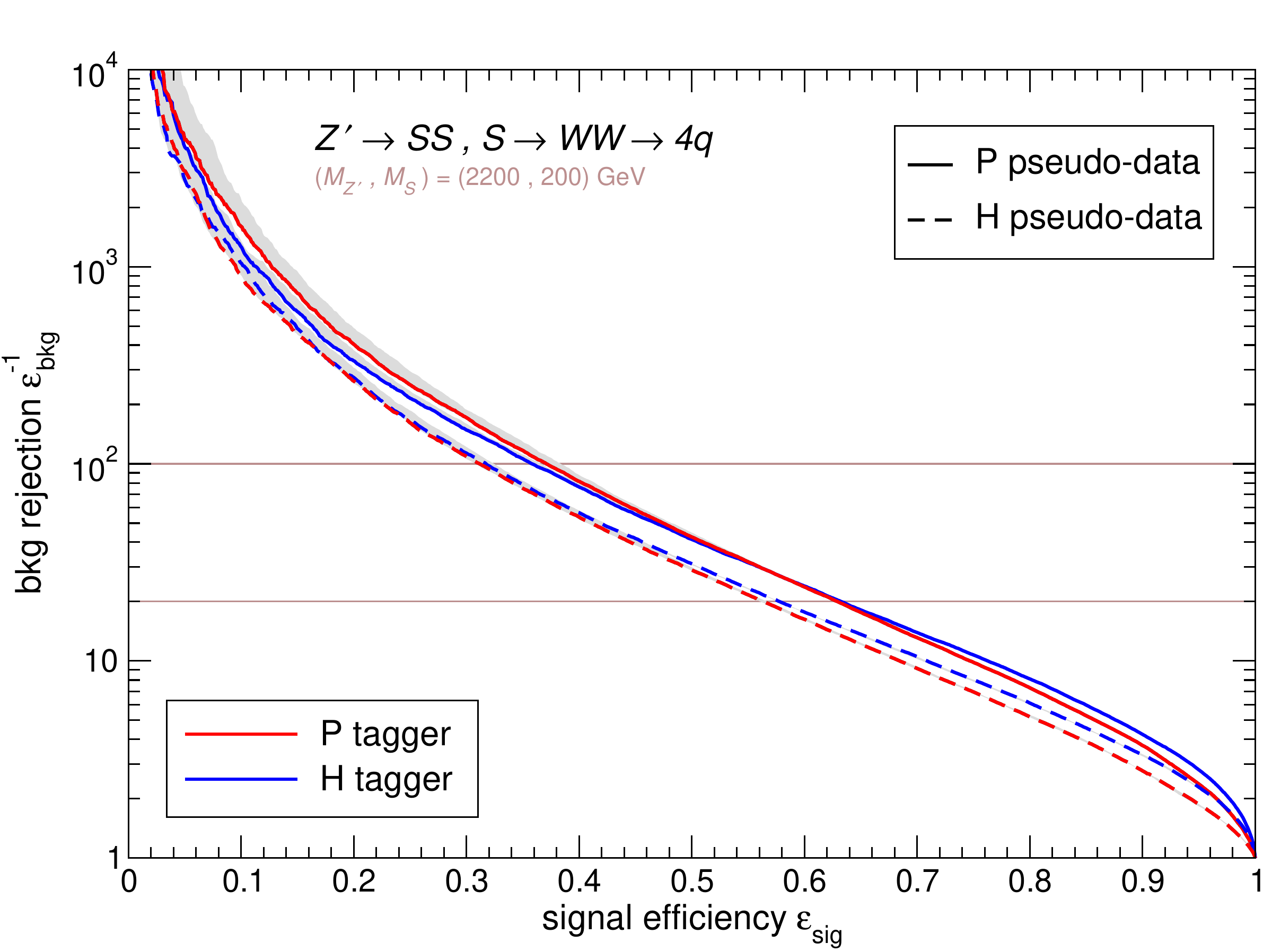}
& \includegraphics[width=8cm,clip=]{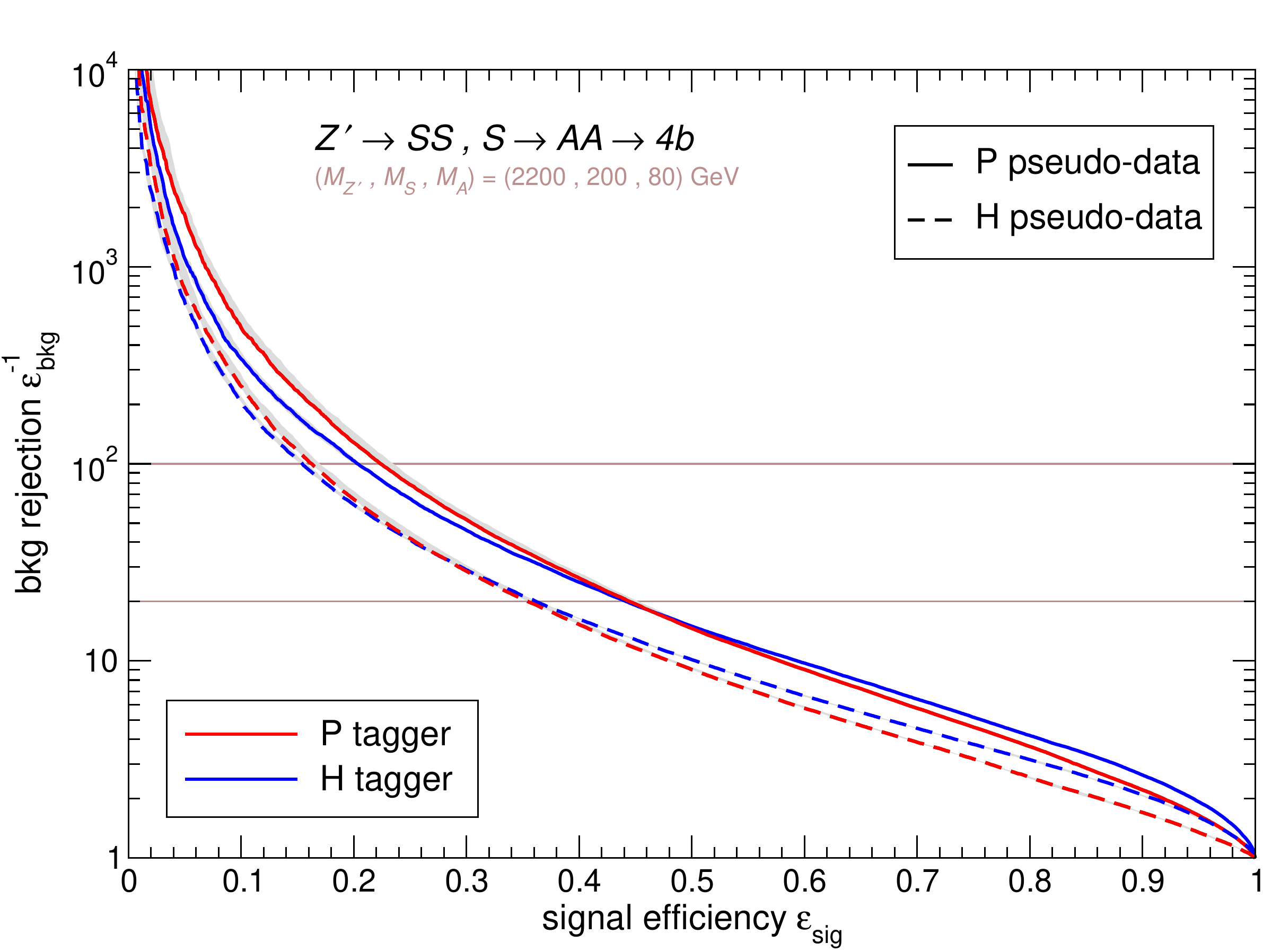}  \\
\includegraphics[width=8cm,clip=]{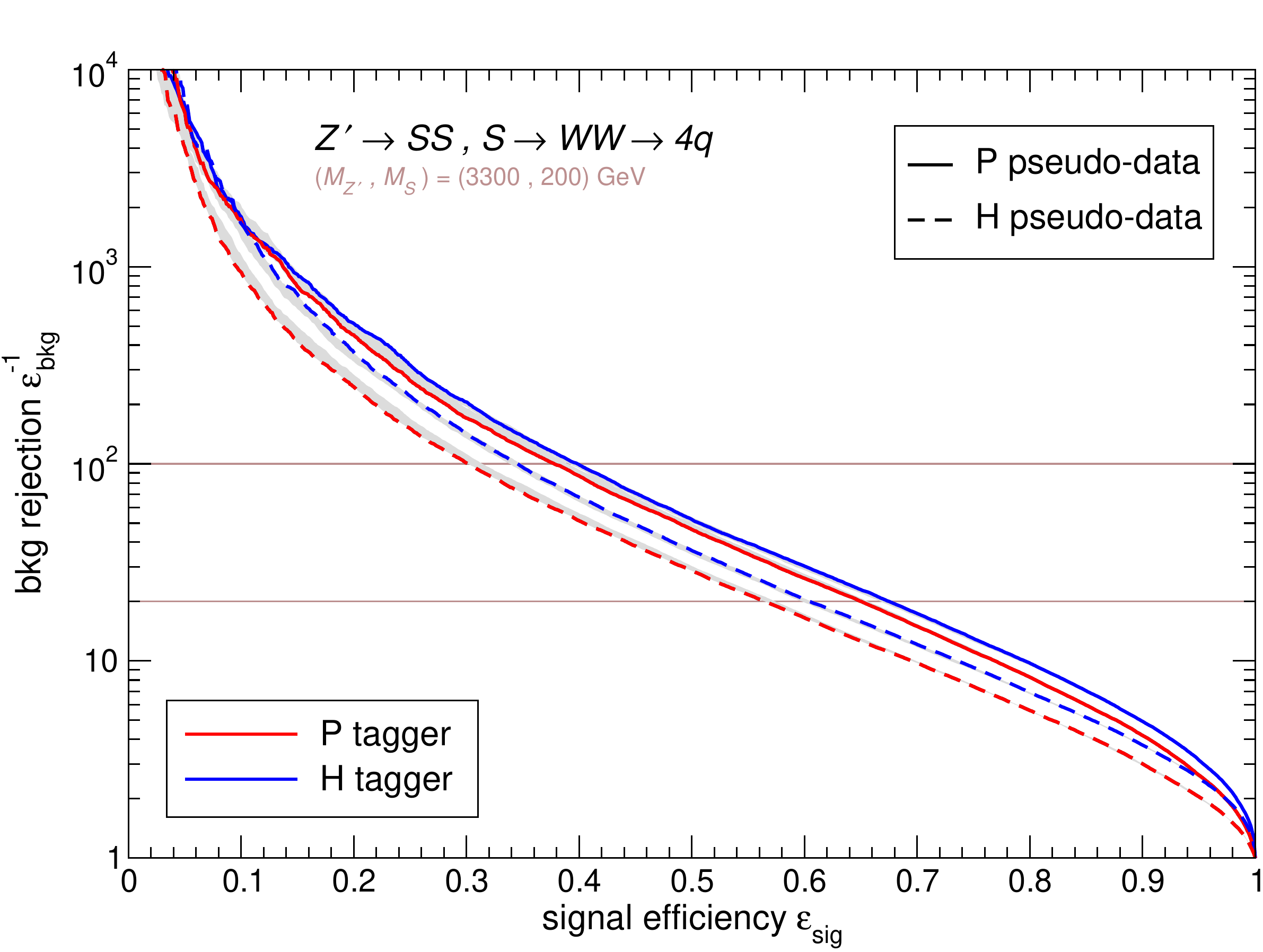}
& \includegraphics[width=8cm,clip=]{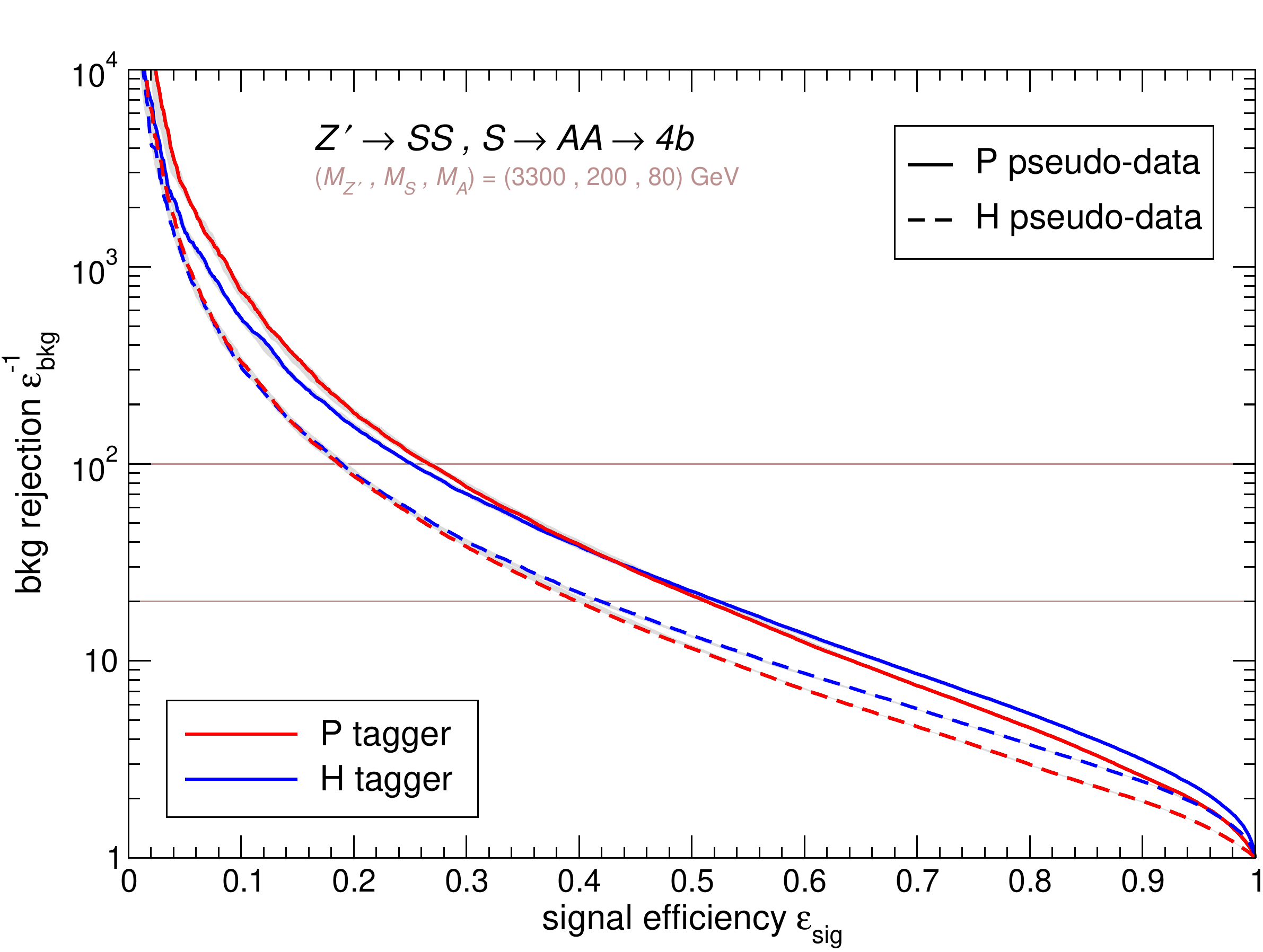} 
\end{tabular}
\caption{ROC curves for 4P jets with $\mj \sim 200$ GeV.}
\label{fig:ROC-4P200}
\end{center}
\end{figure*} 

The $W$ and $S$ benchmarks with $\mj \sim 80$ GeV, $\ptj \sim 1$ TeV were already studied for the anti-QCD tagger~\cite{Aguilar-Saavedra:2017rzt} and the differences between P and H taggers were quite more pronounced than when using MUST. Overall, there are several important conclusions that can be drawn from Figs.~\ref{fig:ROC-2P80} and \ref{fig:ROC-4P80}:
\begin{itemize}
\item[(i)] The P and H pseudo-data have significant differences. This can be seen by comparing, e.g. the two red lines in the plots, which correspond to the same (P) tagger.
\item[(ii)] However, the taggers very effectively learn to discriminate jets of different prongness, independently of the details of the parton shower and hadronisation. The P and H taggers have nearly the same performance on a given pseudo-data set, especially for $W$ bosons. This can be verified by comparing solid and dashed lines of the same colour.
\item[(iii)] Consequently, the ability to distinguish between multi-pronged and QCD jets depends rather on which pseudo-data is considered (in other words,  how pseudo-data is), than on the simulation employed in the tagger training. Pictorially, the curves corresponding to different tagger, same pseudo-data are (much) closer than the curves corresponding to same tagger, different pseudo-data. 
\end{itemize}
The differences between taggers increase with $\ptj$, corresponding to more collimated jets, in which case the higher-order $\tau_n^{(i)}$ are expected to play a more important role in the discrimination.
Also, one can notice that for 2P signals the P tagger is better on P and H pseudo-data, while the opposite behaviour is seen for 4P signals except (partially) for $\ptj \sim 500$ GeV.
We believe this is a consequence of the NN training and the balance in the minimisation of the loss function for several multi-pronged jet MI data, which favours a better discrimination of 2P signals in the case of P training, and a better discrimination of 4P signals in the case of H training. The small spread between trainings (gray band) shows this is not a statistical effect.

Results for 3P top jets are presented in Fig.~\ref{fig:ROC-3P}. With a larger ratio $\mj/\ptj$, the impact of higher-order $\tau_n^{(i)}$ is smaller for the discrimination between signal jets and the background. Consequently, all the ROC curves are quite close for $M_{Z'} = 2.2$ TeV (top panel), and slightly spread for $M_{Z'} = 3.3$ TeV (bottom panel). 

Detailed results for 4P jets of $\mj \sim 200$ GeV, with four light quarks or four $b$ quarks,  are shown in Fig.~\ref{fig:ROC-4P200}. They confirm the claims (i-iii) above. Also as expected, the spread between curves is larger than for 3P jets of similar mass (compare with Fig.~\ref{fig:ROC-3P}) because higher-order $\tau_n^{(i)}$ are more important for the discrimination. For the same reason, the curves are closer for $M_{Z'} = 2.2$ TeV than for $M_{Z'} = 3.3$ TeV, the latter corresponding to more collimated jets.

\begin{figure}[t]
\begin{center}
\begin{tabular}{c}
\includegraphics[width=8cm,clip=]{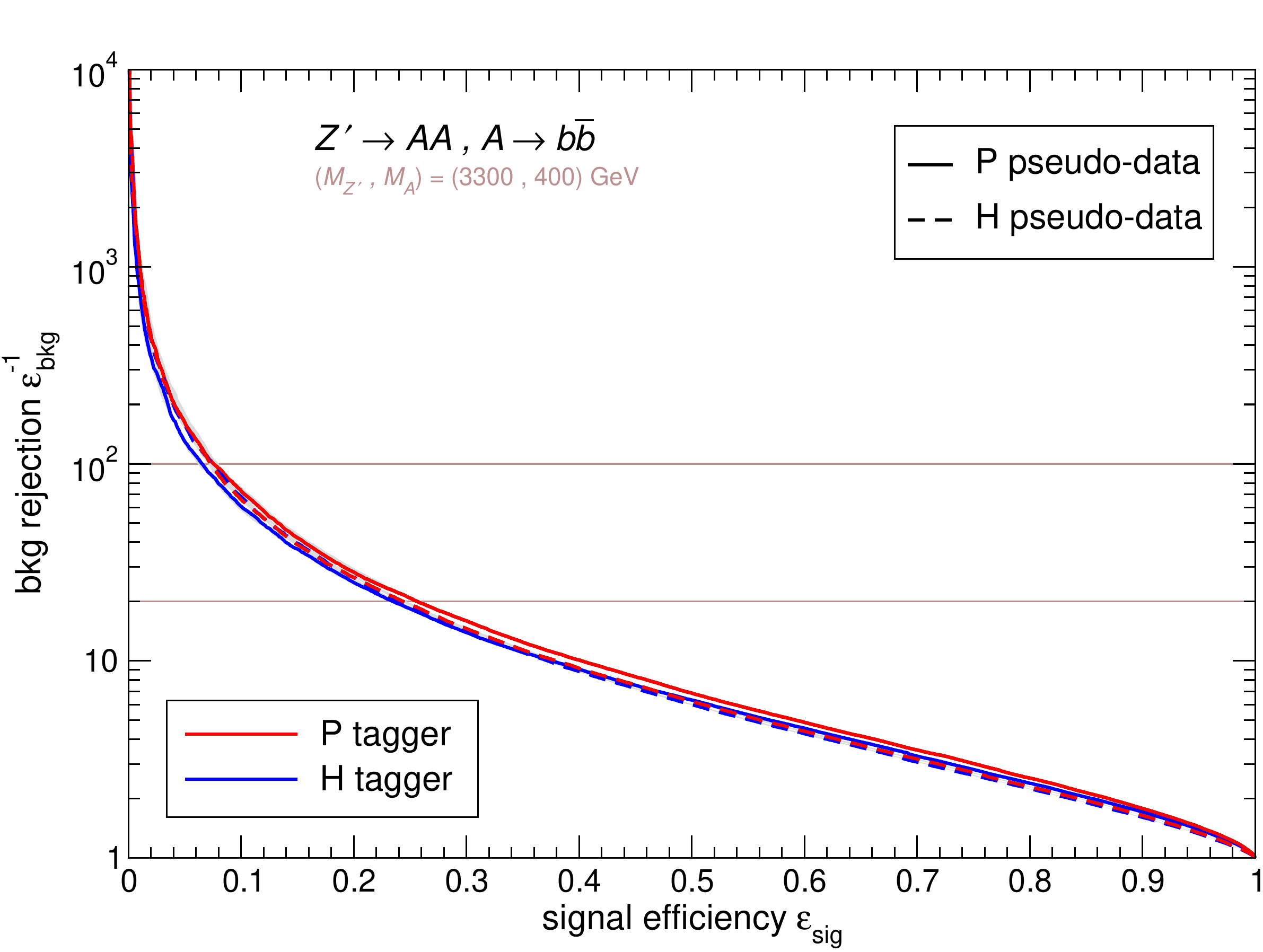} \\
\includegraphics[width=8cm,clip=]{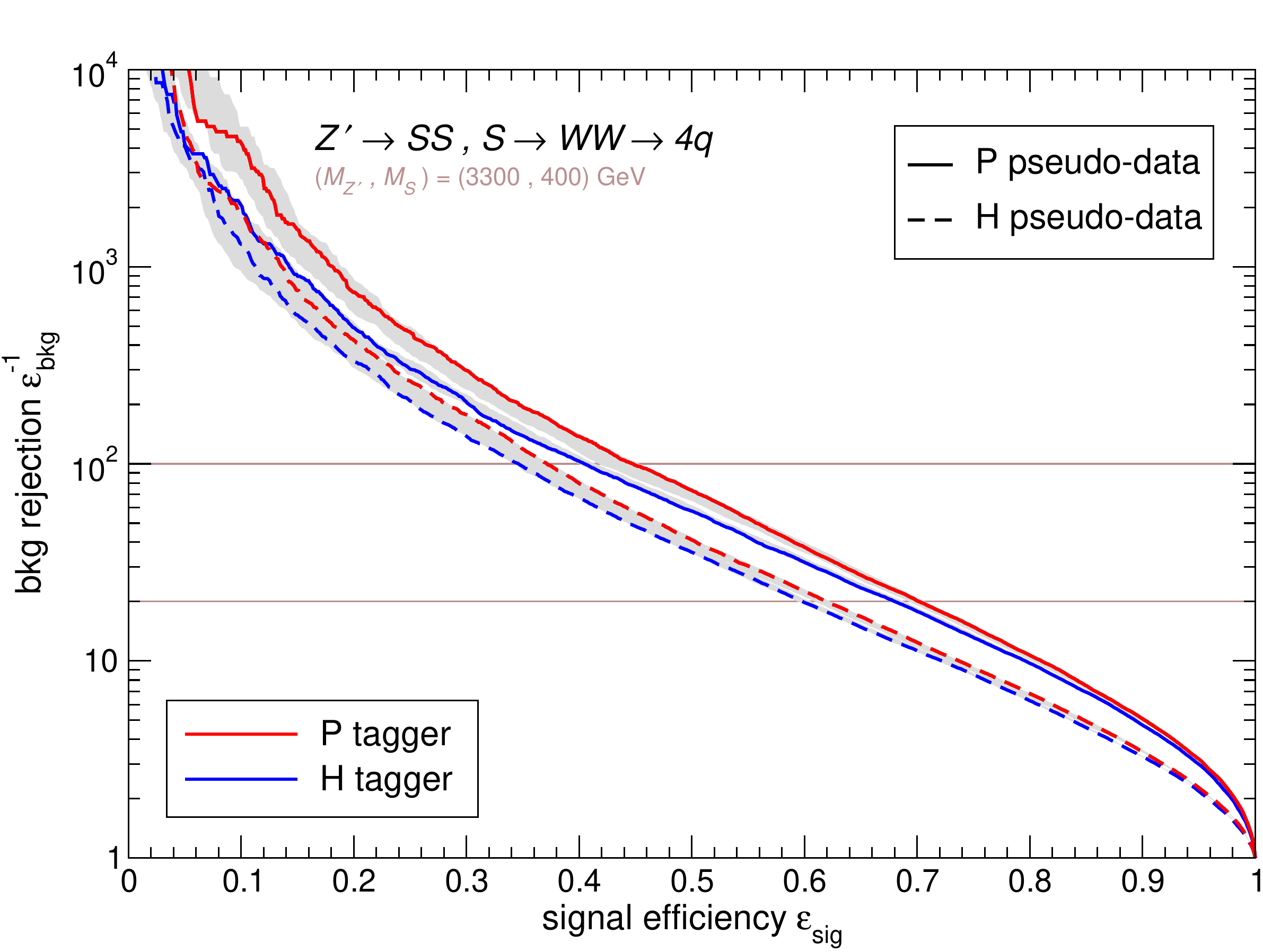} \\ 
\includegraphics[width=8cm,clip=]{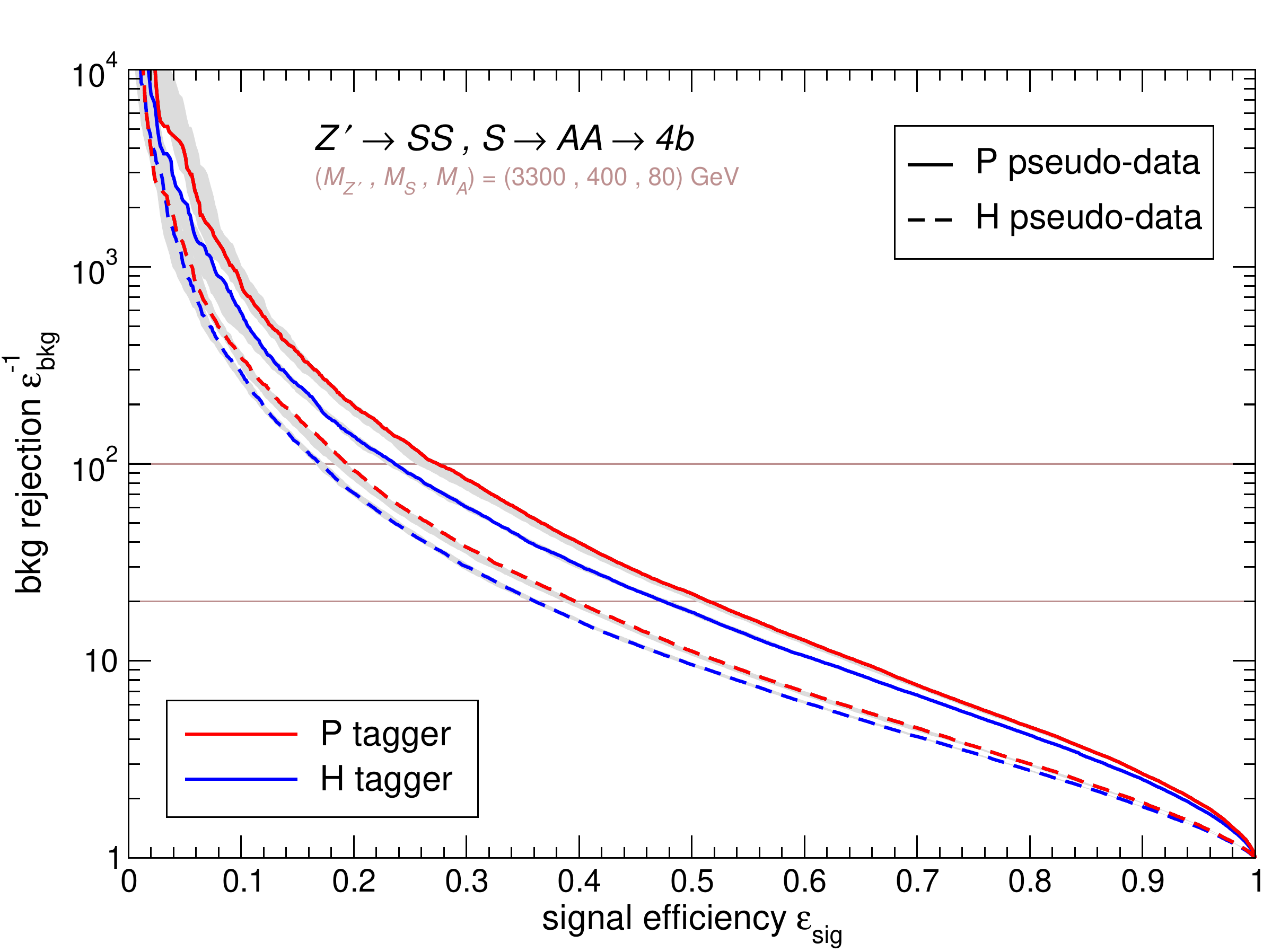} \\ 
\end{tabular}
\caption{ROC curves for jets with $\mj \sim 400$ GeV.}
\label{fig:ROC-M400}
\end{center}
\end{figure}

Finally, we present results for heavier jets with $\mj \sim 400$ GeV in Fig.~\ref{fig:ROC-M400}. Again, the three conclusions (i-iii) above hold, as well as the other two features observed, namely (a) the differences are smaller for 2P than for 4P jets; (b) the curves are closer for larger $\mj/\ptj$. We note that the gray bands are wider in these benchmarks because of the smaller statistics of the samples, which is also seen by the wavy behaviour at low signal efficiencies.

\section{Discussion}
\label{sec:5}

In this paper we have addressed the modeling dependence of jet taggers designed using the MUST method. There are two independent aspects to be considered here: modeling dependence (difference between taggers designed using {\scshape Pythia} or {\scshape Herwig}) and data dependence (difference when a given tagger is applied to either {\scshape Pythia} or {\scshape Herwig} pseudo-data).
From our analysis of 18 benchmarks in section~\ref{sec:4}, two salient conclusions can be drawn:
\begin{itemize}
\item[(i)] Pseudo-data generated with {\scshape Pythia} and {\scshape Herwig} have significant differences.
\item[(ii)] MUST-based taggers very effectively learn to discriminate jets of different prongness, independently of the details of the parton shower and hadronisation. 
\end{itemize}
Here, (i) is inferred from the comparison of same tagger, different pseudo-data, whereas (ii) results from comparing different taggers, and same pseudo-data. Within the several cases analysed, we find that for 2P jets the modeling dependence is insignificant, and completely negligible in some of the benchmarks. For 3P and 4P jets it is small or quite small.

The urging question is, of course, whether either {\scshape Pythia} or {\scshape Herwig} in their different tunes, or other Monte Carlo code for showering and hadronisation like {\scshape Sherpa}~\cite{Sherpa:2019gpd} describe sufficiently well the jet substructure in data, so that supervised generic taggers can be reliably used to search for new physics. Although this cannot be answered only with Monte Carlo studies, the conclusion (ii) above shows that MUST-designed taggers are quite robust and gives confidence in their application to real data. In this regard, possible improvements in the description of the substructure variables of boosted $W$ jets (which can be measured in data) would also benefit the Monte Carlo description for four-pronged jets. 

As discussed above, the variation in the performance of MUST-based taggers is less due to modeling (i.e. whether one uses {\scshape Pythia} or {\scshape Herwig} in the tagger design) and more because of the characteristics of pseudo-data itself, on which the taggers are applied. When applied to real data, the important issue will be whether the subjets within a multi-pronged jet are more or less resolved, so that they are easier or more difficult to distinguish from QCD jets. Of course, this is a feature that also affects unsupervised tools. Consequently, the price to pay by using supervised generic taggers (modeling dependence) may well not be so high, while one can benefit from their better discrimination power.

\section*{Acknowledgements}

I thank  F.R. Joaquim and J. Seabra for previous colaboration in the MUST development.
This work has been supported by grants PID2019-110058GB-C21 and CEX2020-001007-S funded by MCIN/AEI/10.13039/501100011033 and by ERDF, and by FCT project CERN/FIS-PAR/0004/2019.

\end{document}